\documentclass[%
 aip,
rsi,%
 amsmath,amssymb,
 reprint,%
]{revtex4-1}

\usepackage[version=3]{mhchem}
\usepackage[T1]{fontenc} 
\usepackage{amsmath,amssymb,bm}   
\usepackage{cleveref}
\usepackage{xr}
\usepackage{array}
\usepackage{makecell}
\usepackage{graphicx}% Include figure files
\usepackage{dcolumn}% Align table columns on decimal point
\usepackage{bm}% bold math
\begin{document}

\preprint{AIP/123-QED}

\title[Transferabilty of TICs]{On the transferability of time-lagged independent components between similar molecular dynamics systems}

\author{Alexander S. Moffett}
\affiliation{Center for Biophysics and Quantitative Biology, University of Illinois at Urbana-Champaign, Urbana IL, 61801 USA}
\author{Diwakar Shukla}
\affiliation{Center for Biophysics and Quantitative Biology, University of Illinois at Urbana-Champaign, Urbana IL, 61801 USA}
\affiliation{Department of Chemical and Biomolecular Engineering, University of Illinois at Urbana-Champaign, Urbana IL, 61801 USA}
\affiliation{Department of Plant Biology, University of Illinois at Urbana-Champaign, Urbana IL, 61801 USA}
\email{diwakar @ illinois.edu}
\date{\today}

\begin{abstract}
Dimensionality reduction techniques have found great success in a wide range of fields requiring analysis of high-dimensional datasets. Time-lagged independent components analysis (TICA), which finds independent components (TICs) with maximal autocorrelation, is often applied to atomistic biomolecular simulations, where the full molecular configuration can be projected onto only a few TICs describing the slowest modes of motion. Recently, Sultan and Pande have proposed the use of TICs as collective variables for enhanced sampling. However, it is unclear what the best strategy for estimating the TICs of a system is \emph{a priori}. In order to evaluate the utility of TICs calculated on one system to describe the slow dynamics of similar systems, we develop a methodology for measuring the transferability of TICs and apply it to a wide range of systems. We find that transferred TICs can approximate the slowest dynamics of some systems surprisingly well, while failing to transfer between other sets of systems, highlighting the inherent difficulties of predicting TIC transferability. Additionally, we use two dimensional Brownian dynamics simulations on similar potential surfaces to gain insight into the relationship between TIC transferability and potential surface changes.
\end{abstract}
%\keywords{Time Independent Component Analysis; Molecular Dynamics; Transfer Learning; Protein Dynamics}
\maketitle

\section{\label{sec:level1}Introduction}
With the dawn of the ``information age'' has come an overabundance of high-dimensional data, which cannot be properly handled by traditional statistical methods. Dimensionality reduction techniques, such as principal components analysis (PCA), identify the most important directions in a variable space, allowing less important directions to be discarded with minimal loss of information. PCA is a widely used method which accomplishes this goal by finding directions with maximal variance, subject to the constraint that these vectors form an orthogonal, uncorrelated basis set. Though less widely used, time-lagged independent components analysis (TICA)\cite{molgedey1994} has recently been recognized as an important method for finding the slowest collective degrees of freedom (time-lagged independent components, or TICs) in high-dimensional mechanical systems, particularly in simulations of protein conformational dynamics\cite{schwantes2013,perez2013}, which we will focus on in this paper. 

Both PCA and TICA are useful for analyzing datasets \emph{post-hoc}, but can dimensionality reduction methods be used to improve collection of new datasets? Recently, Sultan and Pande performed metadynamics molecular dynamics (MD) simulations using the TIC describing the slowest dynamics of alanine dipeptide as a collective variable, demonstrating the intuitive idea that enhanced sampling on the slowest collective variable of a system leads to rapid convergence of sampling\cite{sultan2017}. Clearly, the slowest degrees of freedom control convergence of sampling and therefore represent prime choices for collective variables in enhanced sampling. However, the proposal to use TICs to guide sampling assumes that the TICs of a system are known \emph{a priori}, requiring unbiased simulation beforehand. While relatively short simulations may be sufficient to estimate the TICs of a small system like alanine dipeptide, far longer simulations may be required for larger systems, negating the improvement in sampling time provided by enhanced sampling.

Instead of performing additional simulations to estimate the TICs of a system, can one use TICs calculated from previous simulations of a similar protein? There is almost certainly not a general answer to this question, as even small changes to the Hamiltonian of an MD system can have large effects on dynamics. However, the use of TICs calculated from simulations of similar proteins, perturbed by a mutation, post-translational modification, or ligand binding, could be a useful heuristic for estimating the slow degrees of freedom of a protein without the need for any additional simulation. This method of applying information gained from sampling one system to improve sampling on another fits under the definition of transfer learning, a concept from machine learning, and accordingly, we will refer to TICs applied to a different system than the one they were calculated on as ``transferred TICs'' from a ``donor system''. We will refer to TICs estimated from the full simulation set of a system as the ``native TICs'' for the ``acceptor system''. 

In a second study, Sultan and Pande proposed this strategy for use in TICA-metadynamics simulations, allowing for efficient sampling of the free energy landscape of alanine dipeptide simulated using different forcefields\cite{sultan2017-2}. However, several questions remain concerning the transfer of TICs between similar systems. More specifically, we would like to answer the following questions concerning transferred TICs:
\begin{enumerate}
\item How does one evaluate the transferability of a set of TICs to a particular system?
\item Is TIC transferability symmetric? That is, if the TICs of one system transfer well to another, do TICs generally transfer in the opposite direction as well?
\item Are there general principles that could determine TIC transferability, and if so what are they?
\end{enumerate}
In order to address these questions, we first develop a mathematical framework to evaluate TIC transferability, by measuring how well transferred TICs reproduce the correlation matrix and the time-lagged autocorrelation matrix of the native TICs of a system. Transferrable TICs will have similar autocorrelations to the target system TICs and will be weakly correlated with one another, meaning that they will approximate the slow dynamics of the target system well and will not provide mutually redundant information. Next, we apply this framework to a diverse set of MD simulations in order to evaluate transferability of TICs between systems differentiated by mutation, ligand binding, and post-translational modification. Finally, we construct simple two-dimensional models representing protein dynamics perturbed in different ways in order to gain insight into how specific changes in the underlying free energy landscape of a protein could determine transferability of TICs.

\section{Evaluating TIC transferability}
TICA takes an $N$-dimensional time series $\mathbf{r}_{t}\in\mathbb{R}^{N}$, with index $t\in\mathbb{N},~t\leq{T}$, and returns $N$ uncorrelated weighting vectors $\mathbf{u}_{i}\in\mathbb{R}^{N}$ (time-lagged independent components, or TICs) with maximal autocorrelation\cite{molgedey1994}. We define the covariance matrix of $\mathbf{r}$ as $\mathbf{C}_{ij}^{r}(0)=(T-1)^{-1}\sum_{t=1}^{T}r_{it}r_{jt}$ and the time-lagged covariance matrix as $\mathbf{C}_{ij}^{r}(\tau)=(T-\tau-1)^{-1}\sum_{t=1}^{T-\tau}r_{it}r_{jt+\tau}$, where $\mathbf{C}^{r}(\tau)$ depends parametrically on the lag time $\tau$. To find the covariance between two vectors $\mathbf{u}_{i}$ and $\mathbf{u}_{j}$, the original time series is projected onto $\mathbf{u}_{i}$ and $\mathbf{u}_{j}$ and the covariance is calculated as $\mathbf{C}^{u}_{ij}(0)=\mathbf{u}_{i}^{\top}\mathbf{C}^{r}(0)\mathbf{u}_{j}$. Similarly, the autocovariance between two vectors is $\mathbf{C}^{u}_{ij}(\tau)=\mathbf{u}_{i}^{\top}\mathbf{C}^{r}(\tau)\mathbf{u}_{j}$. Following P{\'e}rez-Hern{\'a}ndez and coworkers\cite{perez2013}, the slowest overall TIC $\mathbf{u}_{1}$ is first found in the entire $N$-dimensional space by maximizing the Lagrangian $\mathcal{L}(\mathbf{u}_{1})=\mathbf{u}_{1}^{\top}\mathbf{C}^{r}(\tau)\mathbf{u}_{1}-\lambda_{1}\big(\mathbf{u}_{1}^{\top}\mathbf{C}^{r}(0)\mathbf{u}_{1}-1\big)$, in order to maximize autocorrelation of $\mathbf{u}_{1}$ while constraining its variance to one. Next, $\mathbf{u}_{2}$ is found with maximal autocorrelation in a similar manner to $\mathbf{u}_{1}$ but in the subspace satisfying the orthogonality condition $\mathbf{u}_{2}^{\top}\mathbf{C}^{r}(0)\mathbf{u}_{1}=0$. Each subsequent TIC with an index $i$, is found to maximize autocorrelation in the subspace satisfying $\mathbf{u}_{i}^{\top}\mathbf{C}^{r}(0)\mathbf{u}_{j}=0~\forall{j}\in\{1,2,\dots,i-1\}$. 

In summary, the $N$ TICs must satisfy $\mathbf{u}_{i}^{\top}\mathbf{C}^{r}(0)\mathbf{u}_{j}=\delta_{ij}$ and $\mathbf{u}_{i}^{\top}\mathbf{C}^{r}(\tau)\mathbf{u}_{j}=\lambda_{i}\delta_{ij}$ while each $\mathbf{u}_{i}$ maximizes $\lambda_{i}$ in the subspace orthogonal to $\mathbf{u}_{j\in\{1,2,\dots,i-1\}}$. It can readily be shown that TICA solves the generalized eigenvalue problem\cite{perez2013} $\mathbf{C}^{r}(\tau)\mathbf{U}=\bm{\Lambda}\mathbf{C}^{r}(0)\mathbf{U}$ where $\mathbf{U}$ is a matrix with the $N$ TICs in its columns and $\bm{\Lambda}$ is a diagonal matrix of eigenvalues corresponding to autocorrelations.

We are interested in evaluating how well TICs calculated on one system perform on a similar system. Transferrable TICs should have similar autocorrelations to the target system TICs and be weakly correlated with one another in the target system, meaning that they will approximate the slow dynamics of the target system well and will not provide mutually redundant information. In order to test how well a set of transferred TICs, $\mathbf{V}\in\mathbb{R}^{N\times N}$, reproduce the covariance and time-lagged covariance matrices of the native TICs, $\mathbf{U}\in\mathbb{R}^{N\times N}$, of a target system, we calculate the covariance and time-lagged covariance matrices for $\mathbf{V}$ on the target system trajectories, $\mathbf{C}^{v}(0)=\mathbf{V}^{\top}\mathbf{C}^{r}(0)\mathbf{V}$ and $\mathbf{C}^{v}(\tau)=\mathbf{V}^{\top}\mathbf{C}^{r}(\tau)\mathbf{V}$. We then calculate 

\begin{equation} \label{eq:1}
D_{0}\equiv||\mathbf{C}^{v}(0)-\mathbf{I}||_{F}
\end{equation}

\begin{equation} \label{eq:2}
D_{\tau}\equiv||\mathbf{C}^{v}(\tau)-\bm{\Lambda}||_{F}
\end{equation}

where $||\mathbf{M}||_{F}=\sqrt{\sum_{i}\sum_{j}|m_{ij}|^{2}}$ is the Frobenius norm of some matrix $\mathbf{M}$. Both quantities will equal zero if the transferred TICs perfectly estimate the target system TICs.

In practice, a subset of the transferred TICs will be used to estimate the slowest processes of the target system. We would like to determine how well the top $K$ transferred TICs, $\mathbf{V}_{K}\in\mathbb{R}^{N\times{}K}$, can reproduce the top $M\leq K$ target system TICs, $\mathbf{U}_{M}\in\mathbb{R}^{N\times{}M}$. In order to do this, we find the matrix $\mathbf{X}$ which gives the least-squares solution to $\mathbf{V}_{K}\mathbf{X}=\mathbf{U}_{M}$ using the Moore-Penrose pseudoinverse of $\mathbf{V}_{K}$\cite{penrose1956,ben2003}, denoted as $\mathbf{V}_{K}^{+}$, so that $\mathbf{X}=\mathbf{V}_{K}^{+}\mathbf{U}_{M}$. We calculate $\mathbf{V}_{K}^{+}$ using singular value decomposition as implemented in NumPy 1.11\cite{walt2011}. To test the ability of $K$ transferred TICs to represent $M$ native TICs, we then calculate

\begin{equation} \label{eq:3}
D_{KM}\equiv||\mathbf{V}_{K}\mathbf{X}-\mathbf{U}_{M}||_{F}.
\end{equation}

As we are generally most interested in the slowest processes of protein dynamics, we evaluate the performance of the top 2 transferred TICs ($K=2$) in estimating the slowest TIC in the target system ($M=1$) with varying amounts of sampling. We use \cref{eq:1,eq:2,eq:3} to test how much sampling in the target system is needed to match the performance of extended sampling in a similar system in terms of TIC accuracy. 

All quantities described in \cref{eq:1,eq:2,eq:3} are only useful for quantifying relative transferability of TICs. As the idea of transferring TICs is motivated by a desire to limit sampling time, we calculate $D_{0}$, $D_{\tau}$, and $D_{21}$ for truncated donor datasets representing progression of sampling. More specifically, TICA is performed for the donor and acceptor systems truncated to varying degrees and $D_{0}$, $D_{\tau}$, and $D_{21}$ are calculated with respect to trajectories and TICs from the full acceptor system dataset. This approach facilitates comparison between the method of running short unbiased simulations of a system in order to estimate its TICs for use in enhanced sampling and the method of transferring TICs from a similar system while also providing insight into the convergence of TICs to their final values with sampling time.   

In order to clearly denote which systems we are referring to, we use the additional notation $D_{0}(S_{D},S_{A})$, $D_{\tau}(S_{D},S_{A})$, and $D_{21}(S_{D},S_{A})$ for \cref{eq:1,eq:2,eq:3}, meaning that $D_{0}$, $D_{\tau}$, and $D_{21}$ are calculated using the acceptor system simulation set, abbreviated as $S_{A}$, projected onto the transferred TICs from the donor system simulation set, abbreviated as $S_{D}$. For free energy surfaces of one system projected onto the slowest TIC of another system, we use the notation $F(S_{D},S_{A})$ meaning the free energy surface estimated for the acceptor system projected onto the slowest TIC (left implicit for simplicity) of the donor system. Finally, we define relative transfer time as the amount of sampling time needed in an acceptor system to reach the lowest $D_{0}$ value reached by the full donor simulation set, relative to the full acceptor system sampling time. The transfer time is not defined on a donor system for which $D_{0}$ never crosses that of the acceptor system with native TICs.  

\section{Methods}
In this study, we used available simulations from previous studies of the native and GTT mutant FiP35 WW domain\cite{piana2011} (abbreviated as GTT and FiP35 in our notation), cAMP binding domain (cBD) of PKA in cAMP-bound and apo forms\cite{malmstrom2015} (PKA-Holo and PKA-Apo), calmodulin in Ca$^{2+}$-bound and apo forms\cite{shukla2016} (Cal-Holo and Cal-Apo), and BAK1 in glutathionylated and non-glutathionylated forms\cite{moffett2017} (BAK1-C353SG, BAK1-C374SG, BAK1-C408SG, and BAK1-SH). For analysis of the PKA cBD, we only used the longest trajectories initiated from the active state in order to minimize any effects of heterogeneity in sampling method and starting coordinates on our results. Additionally, we performed molecular dynamics simulations on the Met-enkephalin and Leu-enkephalin pentapeptides (Met and Leu).

We constructed Met-enkephalin and Leu-enkephalin as unfolded chains in the tleap program within AmberTools15\cite{case2015} and solvated each with TIP3P water molecules\cite{jorgensen1983} with a buffer of 10 \AA{} between the peptide and the edge of the water box. All energy minimization and simulations were performed in Amber14\cite{case2015} using the Amberff14SB force field\cite{maier2015}. A cutoff of 10 \AA{} was used for non-bonded interactions. For all molecular dynamics simulations, we used an integration time step of 2 fs for Langevin dynamics while constraining all bonds containing hydrogen using the SHAKE algorithm\cite{miyamoto1992}. We treated electrostatic interactions using the particle mesh Ewald method\cite{salomon2013}. We performed all simulations at a constant temperature of 300 K and a constant pressure of 1 atm, maintained using a Langevin thermostat and a Berendsen barostat\cite{berendsen1984}, respectively. We performed 60,000 steps of energy minimization and equilibrated each system for 5 ns. Using the equilibrated structures as starting structures, we ran three independent simulations each of Met-enkephalin and Leu-enkephalin (Table \ref{table:enkeph_sims}).  

Each system was transformed from Cartesian coordinates into the sine and cosine of the $\phi$ and $\psi$ dihedral angles using MSMBuilder 3.8\cite{harrigan2017}. TICA was performed on the transformed trajectories using the MSMBuilder 3.8 implementation with a lag time of 1 frame (see Table \ref{table:lag_times} for lag times in time units). Subsets of simulation sets were taken by ordering simulations in as close to a ``chronological'' order as possible in a given set and taking the first $L$ frames. For example, if a simulation set consisted of 5 independent simulations each with 100 frames, and if we wanted a total of $L=150$ frames, after ordering trajectories we would take all 100 frames of the first trajectory along with the first 50 frames of the second trajectory. We would then perform TICA on the resulting 150 frames, while keeping the 100 frames of the first trajectory distinct from the first 50 frames of the second trajectory.

Gaussian kernel density estimates, as implemented in SciPy 0.18.1 \cite{scipy}, were used to approximate the free energy over the slowest TICs. Free energy was calculated according to $F(x)=-RT\log\left[{p(x)}/{\max_{x}[p(x)]}\right]$, where $R$ is the gas constant in kcal$\cdot$(K$\cdot$mol)$^{-1}$, $T$ is the temperature ($T=300$ K), $p(x)$ is the probability density at $x$, and $\max_{x}[p(x)]$ is the highest point of the probability density function. All plots were created using Matplotlib 1.5.3\cite{hunter2007}.

\section{Transferability of TICs between similar systems}
In order to investigate the transferability of TICs, we evaluate a wide range of protein molecular dynamics simulations from past studies, as well as our own simulations of Met-enkephalin and Leu-enkephalin. We examine how well TICs can be estimated from short simulations of the target system by performing TICA on progressively longer subsets of the available simulation data and measuring $D_{0}$, $D_{\tau}$, and $D_{21}$. The same procedure is repeated using whichever similar systems are to be used for TIC transfer. Finally, we compare projections of each simulation set onto both the slowest native and transferred TICs in an attempt to explain the degree of transferability we observe. As we are interested in the timescales of dynamics along these TICs, we compare the overall shapes of the free energy surfaces.

It is important to emphasize that our analysis focuses on comparison between TICs transferred from a similar system and native TICs estimated with incomplete sampling on the acceptor system, and is not intended to address the success of transferability on an absolute scale. The values $D_{0}$, $D_{\tau}$, and $D_{21}$ are not easily interpretable on their own, but can be used for comparison between transferred TICs and native TICs calculated from truncated trajectories simulating poor sampling or for determining relative transferability of multiple systems to a common acceptor system.
 
\subsection{Leu-enkephalin and Met-enkephalin}
\begin{figure*}[h]
\centering
\includegraphics[width=\textwidth]{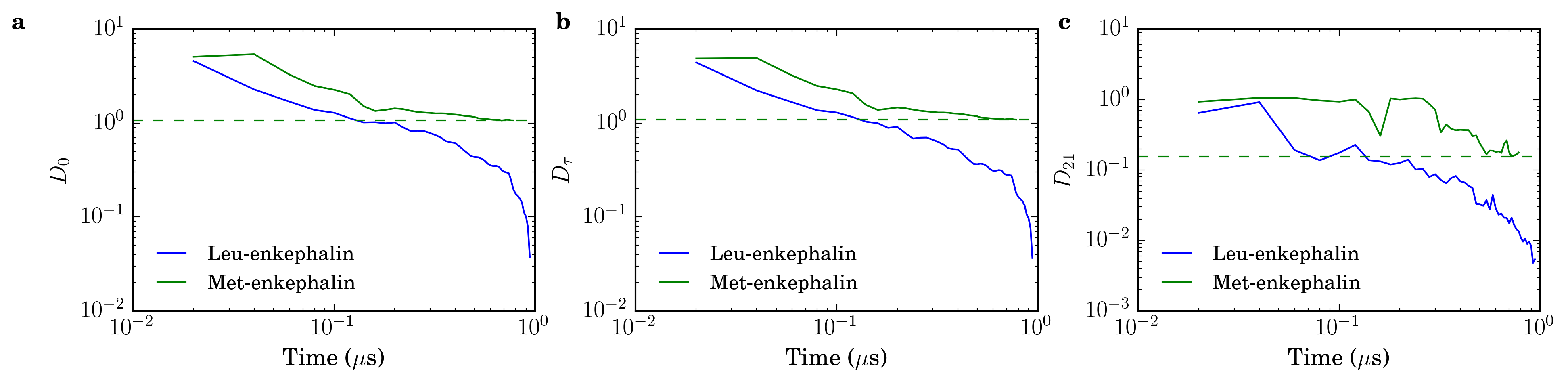}
\caption{Similarity of a) covariance and b) time-lagged covariance matrices for TICs calculated on Met-enkephalin and Leu-enkephalin simulation sets truncated at varying time-lengths on the full Leu-enkephalin simulation set. c) Similarity of the estimated slowest target system (Leu-enkephalin) TIC using the two slowest transferred TICs from Met-enkephalin and Leu-enkephalin simulation sets. The dashed horizontal green lines show the lowest value attained by any Met-enkephalin simulation set.}
\label{fig:on_leu}
\end{figure*}

The full simulation set of Met-enkephalin provides a better approximation to the native TICs of Leu-enkephalin than TICs from simulation sets of Leu-enkephalin truncated up to $\sim0.15~\mu{s}$ (Figure \ref{fig:on_leu}, Table \ref{table:transfer_time}). $D_{0}(Met,Leu)$ and $D_{\tau}(Met,Leu)$ nearly monotonically decrease to their final values, while $D_{21}(Met,Leu)$ behaves slightly more erratically while also ultimately reaching a lower value than $D_{21}(Leu,Leu)$ until $\sim0.15~\mu{s}$ of Leu-enkephalin sampling.  

\begin{figure*}[h]
\centering
\includegraphics[width=\textwidth]{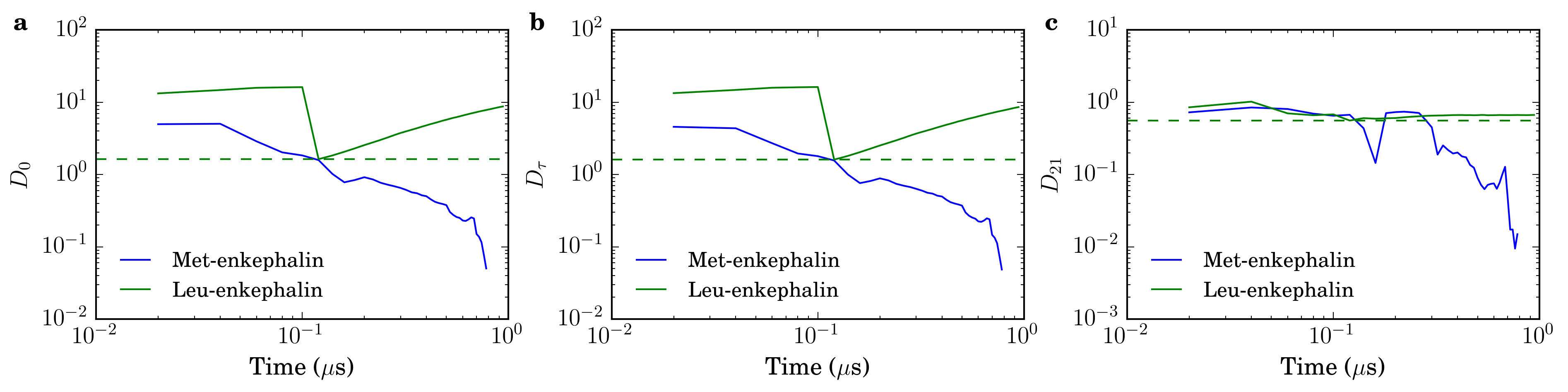}
\caption{Similarity of a) covariance and b) time-lagged covariance matrices for TICs calculated on Leu-enkephalin and Met-enkephalin simulation sets truncated at varying time-lengths on the full Met-enkephalin simulation set. c) Similarity of the estimated slowest target system (Met-enkephalin) TIC using the two slowest transferred TICs from Leu-enkephalin and Met-enkephalin simulation sets. The dashed horizontal green lines show the lowest value attained by any Leu-enkephalin simulation set.}
\label{fig:on_met}
\end{figure*}

Interestingly, there is asymmetry in transferability between Leu-enkephalin and Met-enkephalin, as TICs calculated from Leu-enkephalin simulations appear to poorly approximate Met-enkephalin native TICs overall. Although there is a sharp drop in $D_{0}(Leu,Met)$ and $D_{\tau}(Leu,Met)$ after about $0.1~\mu{s}$, the addition of further sampling raises the values of both above $D_{0}(Met,Met)$ and $D_{\tau}(Met,Met)$ for any amount of sampling (Figure \ref{fig:on_met}a-b). This suggests that the dip in $D_{0}(Leu,Met)$ and $D_{\tau}(Leu,Met)$ near $0.1~\mu{s}$ is likely due to a random fluctuation with an apparently large effect on aggregate protein dynamics amplified by poor sampling. 

\begin{figure*}[h]
\centering
\includegraphics[width=\textwidth]{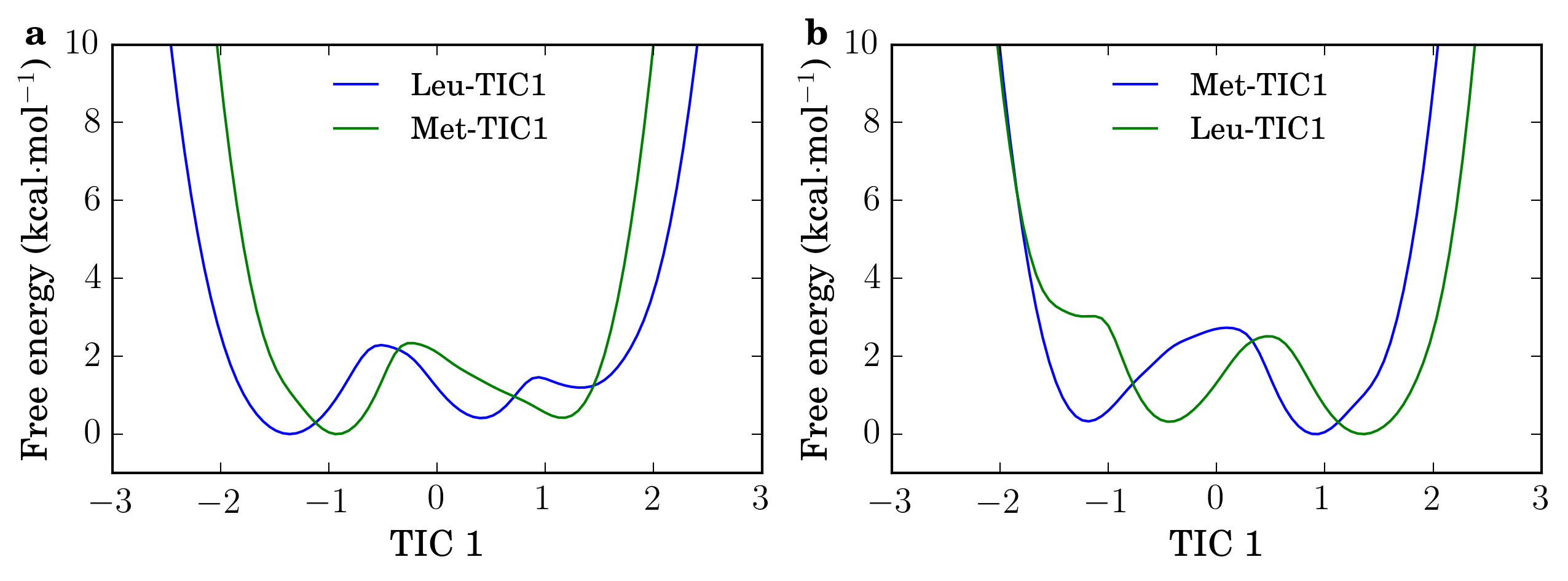}
\caption{Free energy of the a) Leu-enkephalin and the b) Met-enkephalin simulation sets each projected onto both the slowest transferred and native TIC estimated on the full datasets. The x-axis represents a projection of the same data onto a different vector for each plot on the same pair of axes.}
\label{fig:enkeph_prob}
\end{figure*}

In this case, little insight into the cause of this asymmetry can be gained by comparing the free energy surfaces of each system projected onto the slowest TIC the donor system and its own slowest TIC. $F(Leu,Leu)$ and $F(Leu,Met)$ both have a third metastable region not found in $F(Met,Leu)$ or $F(Met,Met)$ which closely resemble one another (Figure \ref{fig:enkeph_prob}). By solely examining these free energy profiles, one might expect the TIC transferability to be higher from Met-enkephalin applied to Leu-enkephalin simulations due to the lower stability of the third peak in $F(Leu,Met)$. This may be accurate at least in terms of $D_{21}$, as the top two Leu-enkephalin TICs approximate the slowest Met-enkephalin TIC well despite the overall failure in transferability of Leu-enkephalin TICs as compared with transfer in the opposite direction (Figure \ref{fig:on_met}c).

\begin{table*}
\begin {center}
\caption{Transfer times between each pair of systems, measured for $D_{0}$.}
\scalebox{0.8}{
\begin{tabular}{ | c c c c c | }
\hline
Donor System & Acceptor System & Transfer Time ($\mu{s}$) & Total Time ($\mu{s}$) & Relative Transfer Time \\ 
\Xhline{4\arrayrulewidth}
Met-enkephalin & Leu-enkephalin & 0.160 & 0.940 & 0.170 \\ 
Leu-enkephalin & Met-enkephalin & 0.120 & 0.780 & 0.154 \\
\hline
GTT mutant & FiP35 WW Domain & 68.800 & 403.200 & 0.171 \\
FiP35 WW Domain & GTT mutant & 16.200 & 620.200 & 0.026 \\
\hline
BAK1-C353SG & BAK1-SH & 5.400 & 34.200 &  0.158 \\ 
BAK1-SH & BAK1-C353SG & 6.600 & 29.800 &  0.221 \\ 
\hline
BAK1-C374SG & BAK1-SH & 5.000 & 34.200 &  0.146 \\ 
BAK1-SH & BAK1-C374SG & 4.000 & 26.600 & 0.150 \\ 
\hline
BAK1-C408SG & BAK1-SH & 0.400 & 34.200 & 0.012 \\ 
BAK1-SH & BAK1-C408SG & 0.200 & 44.800 & 0.004 \\ 
\hline
Apo Calmodulin & Holo Calmodulin & 1.000 & 455.000 & 0.002 \\ 
Holo Calmodulin & Apo Calmodulin & N/A & 256.000 & N/A \\
\hline
Apo PKA cBD & Holo PKA cBD & 8.640 & 16.200 & 0.533 \\
Holo PKA cBD & Apo PKA cBD & N/A & 13.800 & N/A \\ 
\hline
\end{tabular}
}
\label{table:transfer_time}
\end{center} 
\end{table*}

\subsection{Unmodified and S-glutathionylated BAK1 core kinase domain}
\begin{figure*}[h]
\centering
\includegraphics[width=\textwidth]{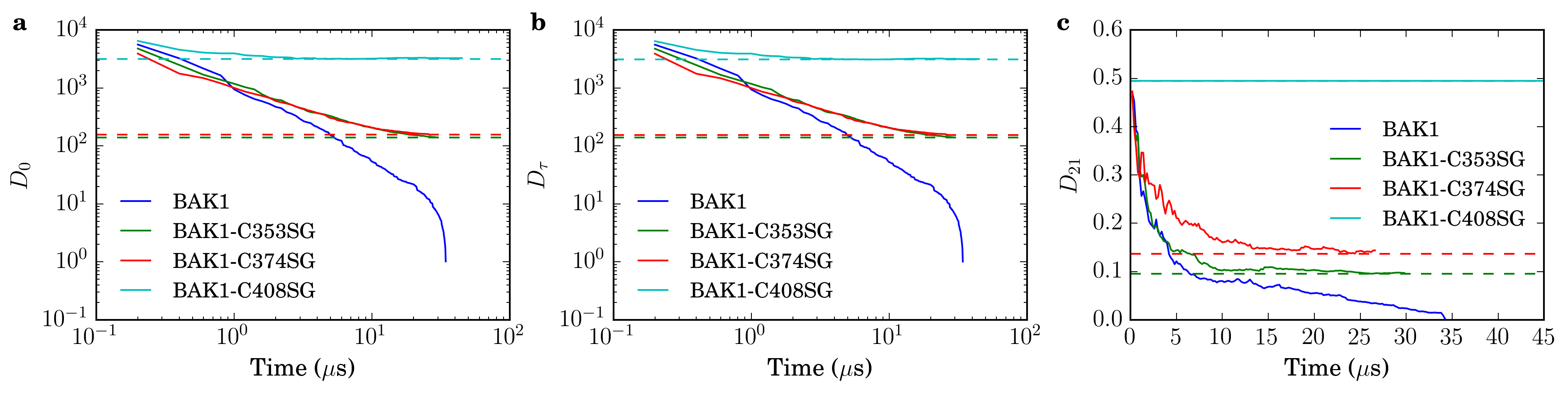}
\caption{Similarity of the a) covariance and b) time-lagged covariance matrices for TICs calculated on BAK1-SH, BAK1-C353SG, BAK1-C374SG, and BAK1-C408SG simulation sets truncated at varying time-lengths on the full BAK1-SH simulation set. Log-log plots are used to improve visual differentiability of different datasets on the same axes. c) Similarity of the estimated slowest target system (BAK1-SH) TIC using the two slowest transferred TICs from BAK1-SH, BAK1-C353SG, BAK1-C374SG, and BAK1-C408SG simulation sets. The dashed horizontal lines show the lowest values attained by any S-glutathionylated BAK1 simulation set.}
\label{fig:bak1_on_atp}
\end{figure*}

TICs estimated on simulations of BAK1 S-glutathionylated on C353 and C374 (referred to as BAK1-C353SG and BAK1-C374SG, respectively) approximate the native TICs of unmodified BAK1 well, ultimately achieving lower $D_{0}$, $D_{\tau}$, and $D_{21}$ values than the native BAK1 TICs for up to $\sim4~\mu$s of BAK1 sampling (Figure \ref{fig:bak1_on_atp}, Table \ref{table:transfer_time}). However, TICs calculated from simulations of BAK1-C408SG poorly approximate native BAK1 TICs as compared with BAK1 TICs calculated with even the smallest amount of BAK1 sampling (Figure \ref{fig:bak1_on_atp}). Increasing sampling for BAK1-C408SG had almost no effect on performance similarity of the transferred TICs on the BAK1 simulations, suggesting a large shift in BAK1 dynamics upon S-glutathionylation of C408. This finding is consistent with the results of the original work from which we have taken these simulations, where S-glutathionylation of C353 and C374 had little effect on BAK1 dynamics while S-glutathionylation on C408 had a significant effect\cite{moffett2017}.

\begin{figure*}[h]
\centering
\includegraphics[width=\textwidth]{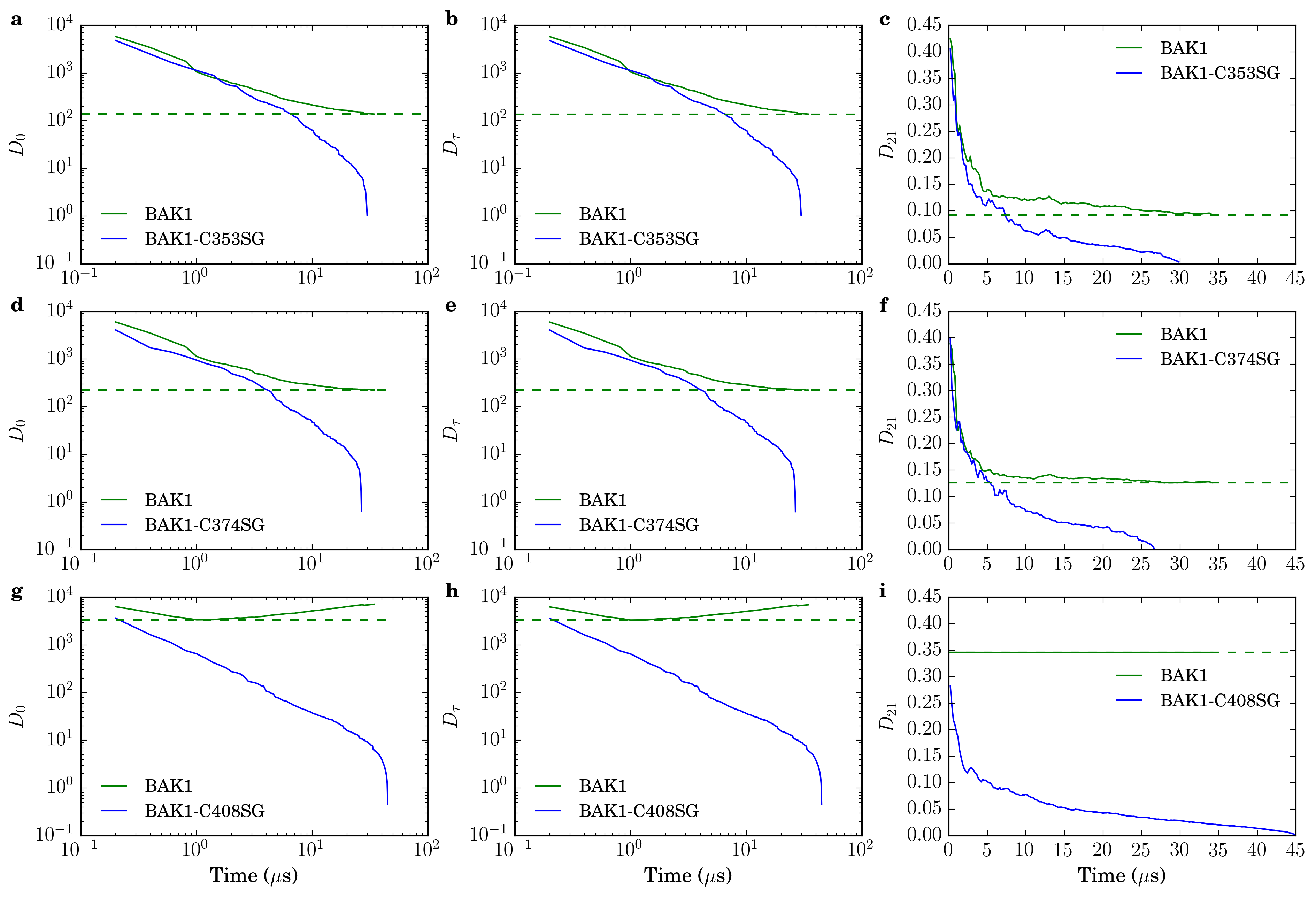}
\caption{Similarity of the a),d),g) covariance and b),e),h) time-lagged covariance matrices for TICs calculated on BAK1-SH, BAK1-C353SG, BAK1-C374SG, and BAK1-C408SG simulation sets truncated at varying time-lengths on the full BAK1-C353SG, BAK1-C374SG, and BAK1-C408SG simulation sets. Log-log plots are used to improve visual differentiability of different datasets on the same axes. c),f),i) Similarity of the estimated slowest target system (BAK1-C353SG, BAK1-C374SG, and BAK1-C408SG, respectively) TIC using the two slowest transferred TICs from BAK1-SH, BAK1-C353SG, BAK1-C374SG, and BAK1-C408SG simulation sets. The dashed horizontal lines show the lowest values attained by any BAK1-SH simulation set.}
\label{fig:bak1_on_gshs}
\end{figure*}

Unlike Leu-enkephalin and Met-enkephalin, unmodified BAK1 and S-glutathionylated BAK1 simulations display symmetric transferability, where TICs from BAK1-SH approximate BAK1-C353SG and BAK1-C374SG TICs well while poorly approximating BAK1-C408SG TICs (Figure \ref{fig:bak1_on_gshs}). 

\begin{figure*}[h]
\centering
\includegraphics[width=\textwidth]{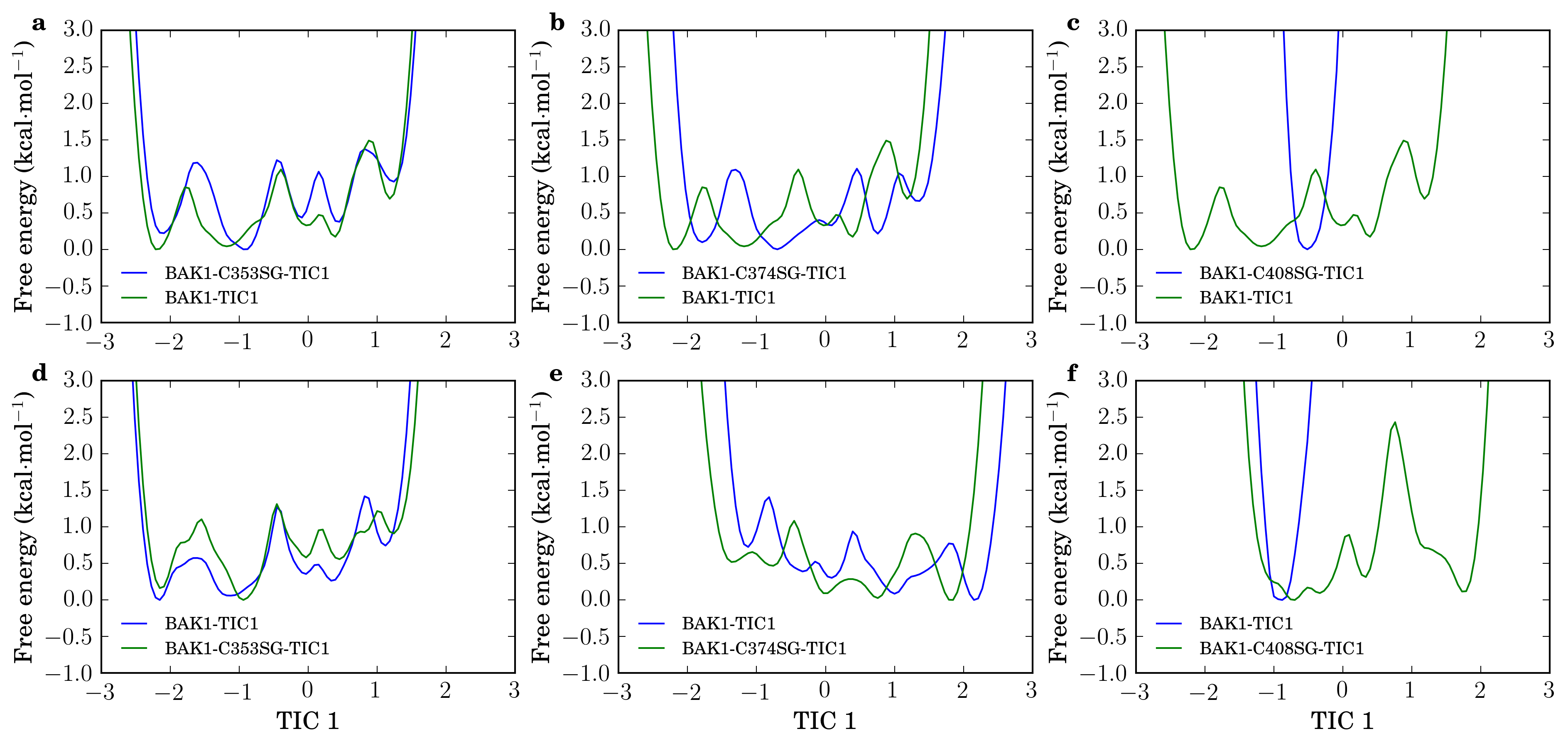}
\caption{Free energy of the a)-c) unmodified BAK1, d) BAK1-C353SG, e) BAK1-C374SG, and f) BAK1-C408SG simulation sets each projected onto both the slowest transferred and native TIC estimated using full datasets. The x-axis represents a projection of the same data onto a different vector for each plot on the same pair of axes.}
\label{fig:bak1_prob}
\end{figure*}

The free energy surfaces of unmodified BAK1 projected onto the slowest BAK1-C353SG and BAK1-C374SG TICs closely resemble the distribution of BAK1-SH projected onto its native slowest TIC (Figure \ref{fig:bak1_prob}a-b). The free energy profiles of BAK1-C353SG and BAK1-C374SG simulations projected onto the slowest BAK1-SH TIC also reproduce the free energy alohe native TICs of each system (Figure \ref{fig:bak1_prob}d-e), explaining the high degree of transferability of TICs between BAK1-SH and the two S-glutathionylated systems. Conversely, $F(BAK1$-$C408SG,BAK1$-$SH)$ and $F(BAK1$-$SH,BAK1$-$C408SG)$ very poorly approximate  $F(BAK1$-$SH,BAK1$-$SH)$ and $F(BAK1$-$C408SG,BAK1$-$C408SG)$, respectively (Figure \ref{fig:bak1_prob}c,f). This result is expected from the poor transferability of TICs between the two systems as measured by $D_{0}$, $D_{\tau}$, and $D_{21}$ and it is clear that the slowest TIC of each system describes a rather fast process in the other, considering the single sharp peaks of both distributions.

\subsection{Apo and Ca$^{2+}$-bound calmodulin}

\begin{figure*}[h]
\centering
\includegraphics[width=\textwidth]{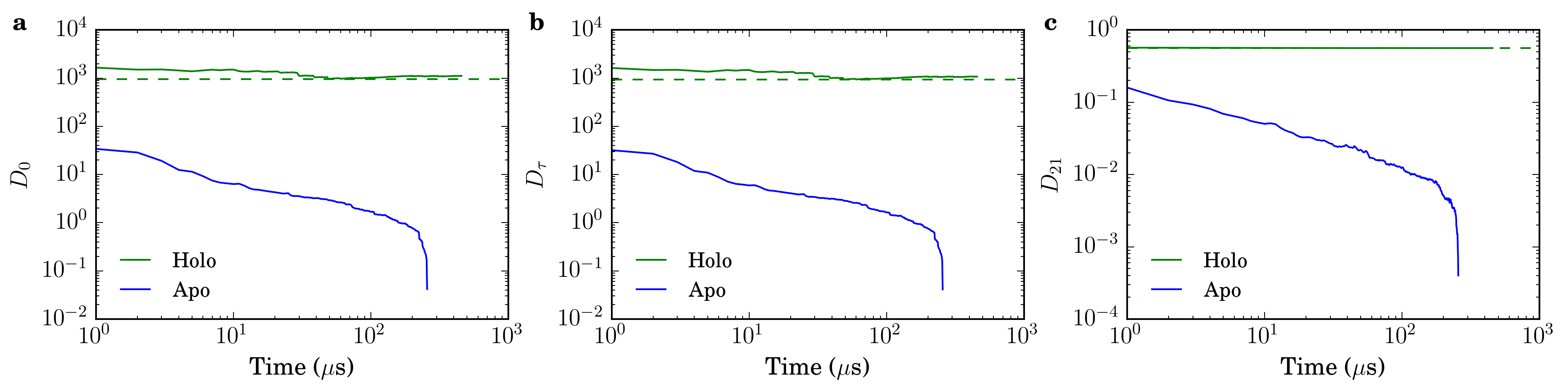}
\caption{Similarity of the a) covariance and b) time-lagged covariance matrices for TICs calculated on apo and holo calmodulin simulation sets truncated at varying time-lengths on the full apo simulation set. Log-log plots are used to improve visual differentiability of different datasets on the same axes. c) Similarity of the estimated slowest target system (apo) TIC using the two slowest transferred TICs from the apo and holo simulation sets. The dashed horizontal green lines show the value attained by any holo calmodulin simulation set.}
\label{fig:cal_on_apo}
\end{figure*}

\begin{figure*}[h]
\centering
\includegraphics[width=\textwidth]{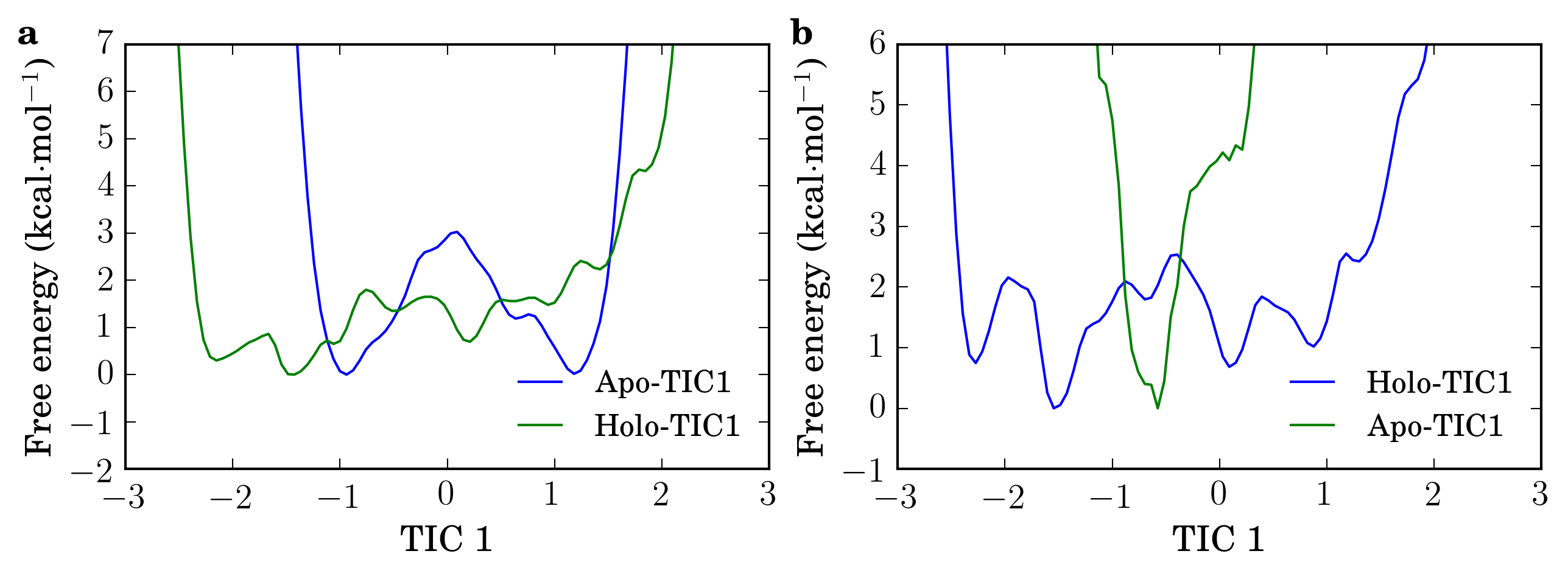}
\caption{Free energy of the a) apo and b) holo calmodulin simulation sets each projected onto both the slowest transferred and native TIC estimated using full datasets. The x-axis represents a projection of the same data onto a different vector for each plot on the same pair of axes.}
\label{fig:cal_prob}
\end{figure*}

The slowest TIC of holo calmodulin performs poorly on apo calmodulin trajectories, as $D_{0}(Holo,Apo)$, $D_{\tau}(Holo,Apo)$, and $D_{21}(Holo,Apo)$ all fail to undercut $D_{0}(Apo,Apo)$, $D_{\tau}(Apo,Apo)$, and $D_{21}(Apo,Apo)$ at any amount of sampling (Figure \ref{fig:cal_on_apo}). While $D_{0}(Apo,Holo)$ and $D_{\tau}(Apo,Holo)$ also largely remain above $D_{0}(Holo,Holo)$ and $D_{\tau}(Holo,Holo)$, the slowest apo calmodulin TIC calculated using the whole dataset provides a decent approximation for the native holo calmodulin TIC calculated with a small amount of sampling relative to the full sampling time of either system (Figure \ref{fig:cal_on_apo}, Table \ref{table:transfer_time}). It is surprising that apo TICs would approximate holo TICs rather than the other way around, as $F(Holo,Apo)$ and $F(Apo,Apo)$ are far more similar to one another than $F(Apo,Holo)$ and $F(Holo,Holo)$ are, where $F(Apo,Holo)$ clearly describes a relatively fast process with no internal barriers (Figure \ref{fig:cal_prob}). In this case, it is clear that there are other factors contributing to TIC transferability than the similarity of free energy landscapes on slowest TICs.

\begin{figure*}[h]
\centering
\includegraphics[width=\textwidth]{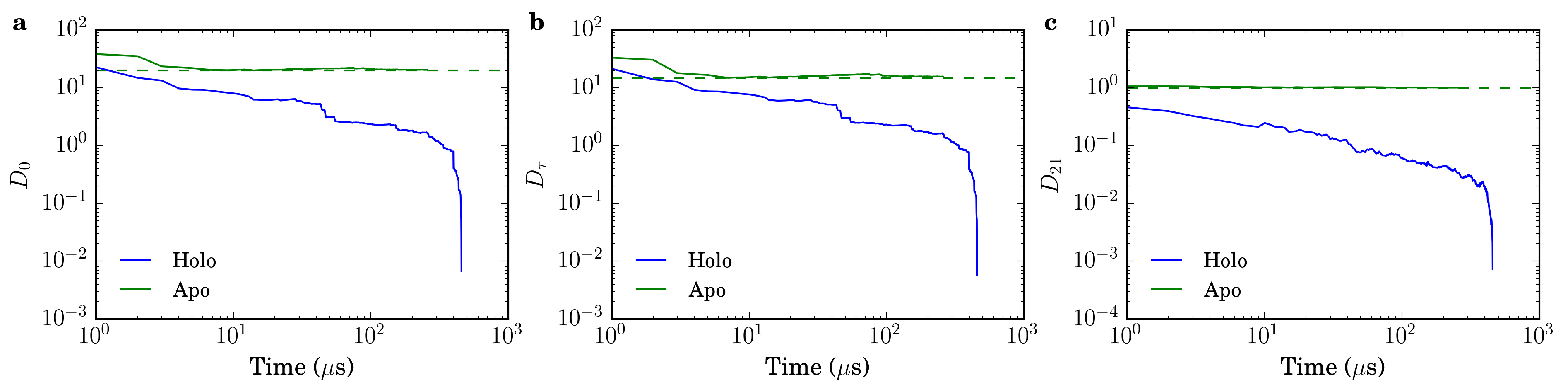}
\caption{Similarity of the a) covariance and b) time-lagged covariance matrices for TICs calculated on apo and holo calmodulin simulation sets truncated at varying time-lengths on the full holo simulation set. Log-log plots are used to improve visual differentiability of different datasets on the same axes. c) Similarity of the estimated slowest target system (holo) TIC using the two slowest transferred TICs from the apo and holo simulation sets. The dashed horizontal green lines show the value attained by any apo calmodulin simulation set.}
\label{fig:cal_on_holo}
\end{figure*}

\subsection{FiP35 WW domain and GTT mutant}
The FiP35 WW domain and its GTT triple mutant display a high degree of mutual TIC transferability, possibly explaining the success of transferrable TICA-metadynamics simulations, where wild-type FiP35 structures were swapped into mutant simulations according to an acceptance criterion\cite{sultan2017-2}.  TICs transferred from the GTT mutant to the wild-type FiP35 WW domain yield lower $D_{0}$ and $D_{\tau}$ values than the native TICs calculated on up to $\sim68~\mu$s of FiP35 sampling (Figures \ref{fig:fip35_on_native}a-b \& \ref{fig:fip35_on_gtt}a-b, Table \ref{table:transfer_time}), while values of $D_{21}(GTT, FiP35)$ and $D_{21}(FiP35, GTT)$ remain below $D_{21}(FiP35, FiP35)$ and $D_{21}(GTT, GTT)$, respectively, for up to $\sim16~\mu$s of native system sampling (Table \ref{table:transfer_time}). Corresponding with the high degree of TIC transferability, the free energy landscapes for both systems along their native slowest TIC closely resemble the free energy landscape along the slowest transferred TIC from the other system (Figures \ref{fig:fip35_prob}c \& \ref{fig:fip35_prob}).

\subsection{Apo and cAMP-bound PKA cAMP-binding domain}
The covariance and time-lagged covariance matrices of the PKA cBD behave erratically, where $D_{0}$ and $D_{\tau}$ drastically increase with the first microsecond of sampling (Figures \ref{fig:pka_on_apo}a-b \& \ref{fig:pka_on_holo}a-b). Use of the slowest holo TIC provides little advantage over the native apo slowest TIC, while the slowest native holo TIC cannot be well approximated using the two slowest TICs calculated with truncated sampling of the apo or holo systems (Figure \ref{fig:pka_on_apo}). Strangely, $D_{0}(Apo,Holo)$ and $D_{\tau}(Apo,Holo)$ are lower than $D_{0}(Holo,Holo)$ and $D_{\tau}(Holo,Holo)$ when compared at the same truncated sampling times up until $8~\mu$s of sampling (Figure \ref{fig:pka_on_holo}). While this behavior is unexpected and could indicate some issue with the dataset or our handling of the analysis, it appears that the slowest apo TIC is transferrable to the holo system, although $D_{21}(Apo,Holo)$ fails to decrease with increased apo sampling (Figure \ref{fig:pka_on_holo}c). 

$F(Holo,Apo)$ clearly explains the lack of transferability of the holo TICs to the apo system, as it fails to cover nearly half of the TIC covered by a low free energy region of $F(Apo,Apo)$ (Figure \ref{fig:pka_prob}a). The apparently higher degree of transferability of the apo TICs to the holo system is not explained by the markedly different $F(Apo,Holo)$ and $F(Holo,Holo)$ (Figure \ref{fig:pka_prob}b), as $F(Apo,Holo)$ clearly describes a different process which is also likely slow due to the large interior free energy barrier.

\subsection{Which changes to the free energy landscape permit transferability?}
\begin{figure*}[h]
\centering
\includegraphics[width=\textwidth]{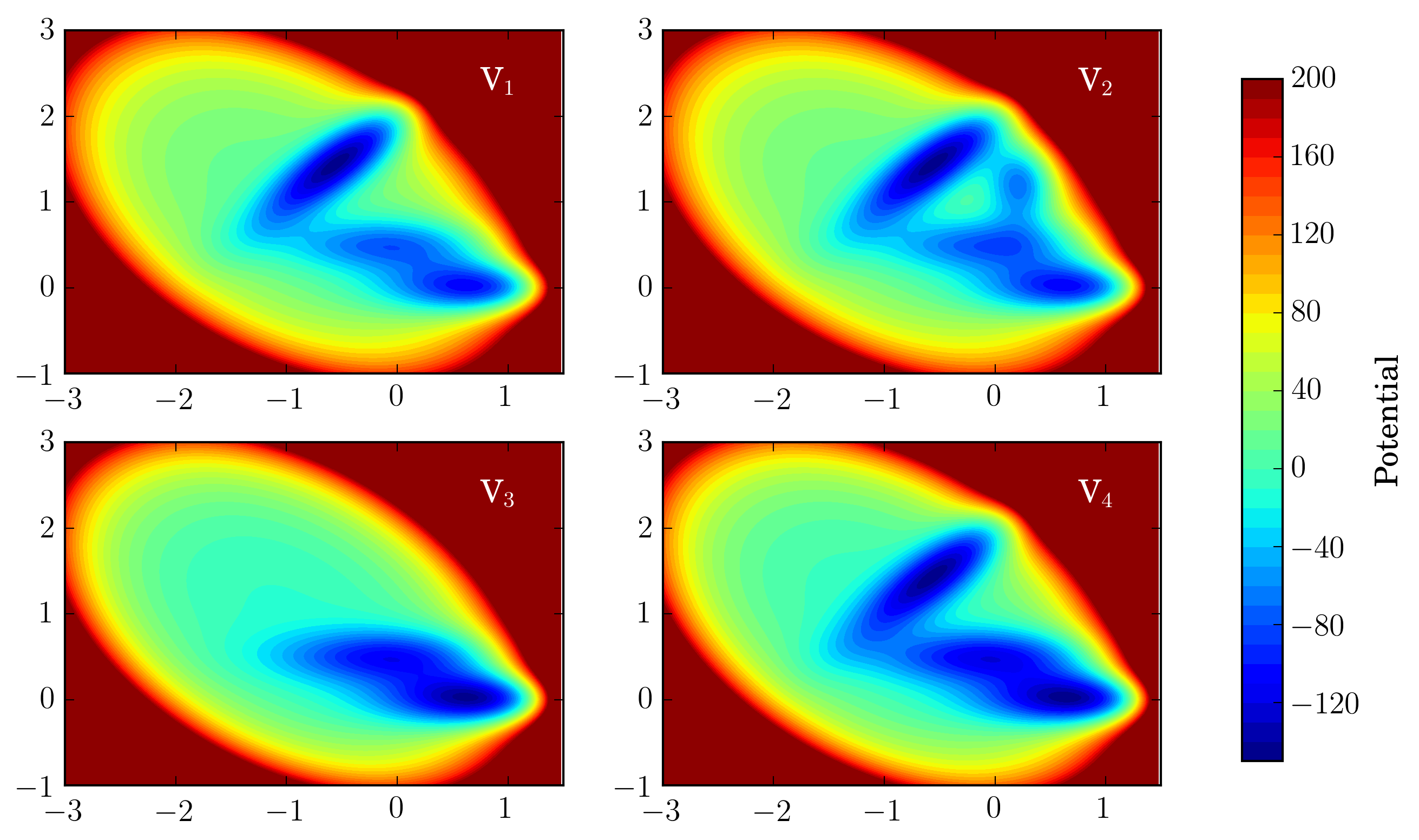}
\caption{The four potential surfaces used for Brownian dynamics simulations. All units in the Brownian dynamics simulations were arbitrary, as any connection to physical units was unnecessary for illustrative purposes.}
\label{fig:bd_potentials}
\end{figure*}

From testing TIC transferability for a variety of systems, it is clear that TIC transferability is highly system-dependent. In order to illustrate which differences between two free energy landscapes could allow for TIC transferability, we performed Brownian dynamics simulations of a single particle on a two-dimensional M{\"u}ller-like potential (V$_{1}$) as well as several alternate potentials, one with a basin of attraction added (V$_{2}$), another with a basin of attraction removed (V$_{3}$), and a third with the same number of basins but with altered relative energies (V$_{4}$) (Figure \ref{fig:bd_potentials}). We then applied our measures of TIC transferability to these simulations. Although these simulations represent greatly simplified models of protein dynamics and only capture a small number of the possible ways a free energy landscape could be altered with some perturbation, they serve to illustrate the connection between TIC transferability and possible underlying causes.

\begin{figure*}[h]
\centering
\includegraphics[width=\textwidth]{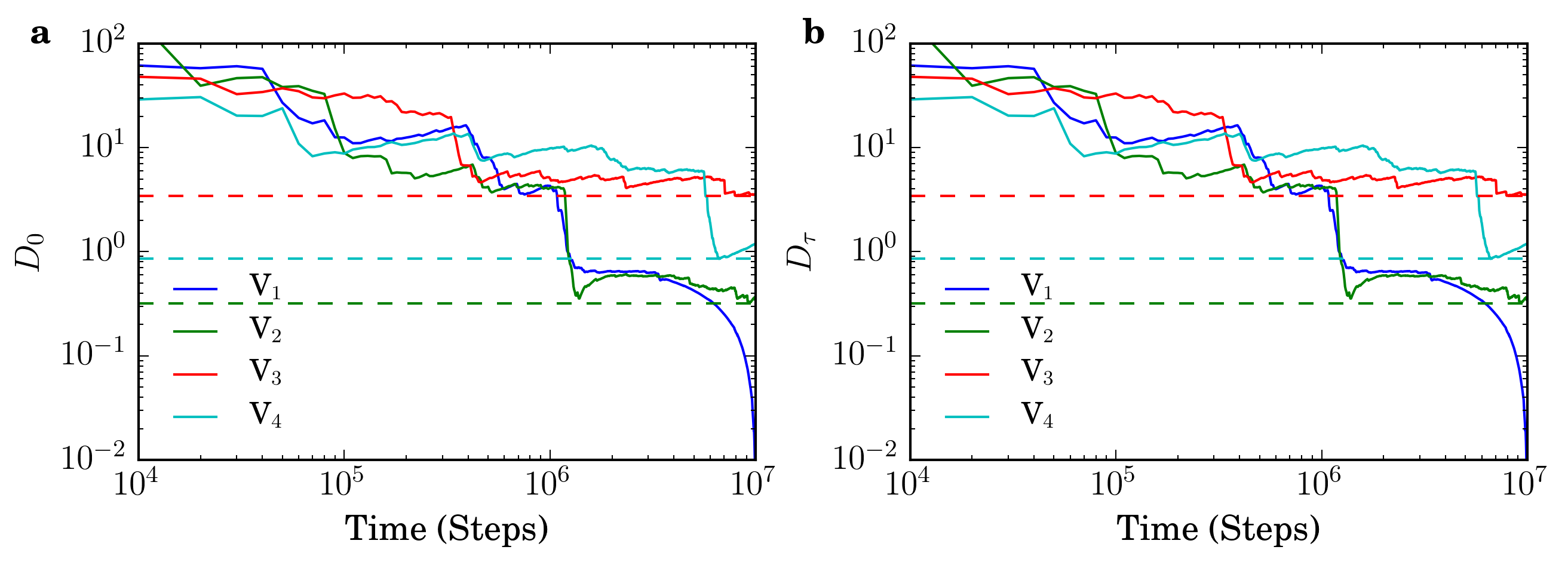}
\caption{Similarity of the a) covariance and b) time-lagged covariance matrices for TICs calculated on the V$_{1}$-V$_{4}$ simulation sets truncated at varying time-lengths on the full V$_{1}$ simulation set. Log-log plots are used to improve visual differentiability of different datasets on the same axes. The dashed horizontal lines show the lowest value attained by any V$_{2}$-V$_{4}$ simulation set.}
\label{fig:bd_on_v1}
\end{figure*}

TICs calculated from simulations on V$_{2}$ appear to approximate V$_{1}$ TICs best, as $D_{0}(V_{2},V_{1})$ and $D_{\tau}(V_{2},V_{1})$ closely follow $D_{0}(V_{1},V_{1})$ and $D_{\tau}(V_{1},V_{1})$, while unsurprisingly V$_{4}$ performs worst of the three alternate potentials (Figure \ref{fig:bd_on_v1}). Intuitively, sampling on all three alternate potentials produce similar $D_{0}$ and $D_{\tau}$ values to sampling on V$_{1}$ up to $\sim{}10^{6}$ steps of sampling, when the particles on V$_{1}$ and V$_{2}$ transition into the upper-left potential well, exploring a new mode of motion (Figure \ref{fig:v1_movie} \& \ref{fig:v2_movie}). At this point, $C_{r}(0)$ and $C_{r}(\tau)$ of sampling on V$_{3}$ and all other systems permanently diverge, as V$_{3}$ lacks the upper-left well, while sampling on V$_{4}$ eventually reaches the upper-left well at $\sim{}6\cdot{}10^{6}$ steps (Figure \ref{fig:v4_movie}), resulting in a rapid decrease in $D_{0}(V_{4},V_{1})$ and $D_{\tau}(V_{4},V_{1})$ (Figure \ref{fig:bd_on_v1}b).

Similarly, TICs from V$_{1}$ are highly transferable to V$_{2}$ and V$_{4}$ simulations, as well as V$_{3}$ simulations up until the jump into the upper-left basin near $10^{6}$ steps (Figure \ref{fig:bd_on_v2-4}). We do not observe any asymmetry in TIC transferability between any of the examined two-dimensional potentials.

It is clear that removal of a large metastable potential well, even with an otherwise unmodified potential surface, can significantly decrease transferability of TICs between two systems, as seen with Brownian dynamics simulations on V$_{1}$ and V$_{3}$. However, addition of a metastable well near already existing pathways, as with V$_{2}$ compared with V$_{1}$, appears to have little effect on TIC transferability. Minor alterations of relative well stability, as in V$_{4}$, also appear to allow for TIC transferability. It is easy to imagine exceptions to these observations; for example, a perturbation which removes a state of little kinetic relevance would likely provide a system amenable to TIC transfer. This highlights the system specificity of TIC transferability, indicating that caution is warranted when using methods relying on the transfer of TICs between systems.

\section{Conclusions}
We have found that TICs transferred between similar systems can sometimes better approximate native TICs than TICs calculated from short simulations, even for large and complex systems. However, this attribute is, unsurprisingly, highly system-dependent and therefore difficult to predict \emph{a priori}. For example, the apparent asymmetry in transferability between Leu-enkephalin and Met-enkephalin would be difficult to predict without extensive simulation of both, even though one might expect their slow dynamics to closely resemble one another. This unpredictability could be a major issue in the overall applicability of direct transfer of TICs between related systems, highlighting the need for further development of transfer learning-based approaches to accelerating learning of protein dynamics, in the spirit of the structural reservoir of wild-type FiP35 conformations used in Ref. \citenum{sultan2017-2} to accelerate GTT mutant sampling. 

Brownian dynamics simulations on toy potentials allow us to gain some insight into how changes in a potential or free energy landscape can affect transferability of TICs as well as the origins of $D_{0}$ and $D_{\tau}$ behavior. While the four potentials we have examined are highly simplified models of protein dynamics and represent a small subset of the possible changes in an energy landscape with a perturbation, it is clear that small changes along the main pathways of dynamics can largely preserve the TICs of a system (V$_{2}$ and V$_{4}$), while the complete disappearance of a metastable state with a large contribution to slow dynamics will likely decrease TIC transferability (V$_{3}$).

It is clear that the congruity in convergence of $D_{0}$ and $D_{\tau}$ between V$_{1}$ and V$_{2}$ (Figures \ref{fig:bd_on_v1} \& \ref{fig:bd_on_v2-4}a-b) arises from transitions in both systems around the same time into the deep upper left-most potential energy well (Figures \ref{fig:v1_movie} \& \ref{fig:v2_movie}). Even though V$_{4}$ has the same potential energy well, the transition occurs long after those of V$_{1}$ and V$_{2}$ (Figure \ref{fig:v4_movie}), leading to a delay in the sudden $D_{0}$ and $D_{\tau}$ drop observed in V$_{1}$ and V$_{2}$ simulations (Figure \ref{fig:bd_on_v1}). This particular example highlights the role of stochasticity in the transferability of TICs, as a chance fluctuation in the simulated dynamics of a protein could lead to an artificially high or low calculated degree of TIC transferability between systems, provided only limited sampling is performed.\\

\begin{acknowledgments}
The authors would like to thank D. E. Shaw Research for providing the native and GTT mutant FiP35 WW domain trajectories, as well as the research group of Rommie Amaro for making the PKA cBD trajectories available. The authors would also like to thank the Blue Waters sustained-petascale computing project, which is supported by the National Science Foundation (awards OCI-0725070 and ACI-1238993) and the state of Illinois, for providing computing time for this study.
\end{acknowledgments}
%\nocite{*}
%\bibliography{tic_transfer}
%
\clearpage
\newpage

\begin{widetext}
\textbf{Supporting Information: On the transferability of time-lagged independent components between similar molecular dynamics systems}

\renewcommand*{\thepage}{S\arabic{page}}
\renewcommand{\thefigure}{S\arabic{figure}}
\renewcommand{\thetable}{S\arabic{table}}
\setcounter{page}{1}
\setcounter{figure}{0}
\setcounter{table}{0}
\begin{table*}[h]
\begin {center}
\caption{Simulation times for Met-enkephalin and Leu-enkephalin.}
\scalebox{0.8}{
\begin{tabular}{ | c | c | c | c | }
\hline
System & Replicate 1 (ns) & Replicate 2 (ns) & Replicate 3 (ns) \\ 
\Xhline{4\arrayrulewidth}
Met-enkephalin & 319.24 & 371.12 & 319.56 \\ 
\hline
Leu-enkephalin & 300.84 & 320.96 & 320.98 \\
\hline
\end{tabular}
}
\label{table:enkeph_sims}
\end{center} 
\end{table*}

\begin{table*}[h]
\begin {center}
\caption{TICA lag times for all systems.}
\scalebox{0.8}{
\begin{tabular}{ | c | c | }
\hline
System & Lag Time (ps) \\ 
\Xhline{4\arrayrulewidth}
BAK1 & 200 \\ 
\hline
Calmodulin & 100 \\
\hline
Enkephalins & 200 \\
\hline
FiP35 & $\sim$200 \\ 
\hline
PKA cBD & 120 \\ 
\hline
\end{tabular}
}
\label{table:lag_times}
\end{center} 
\end{table*}

\newpage

\begin{figure*}[h]
\centering
\includegraphics[width=\textwidth]{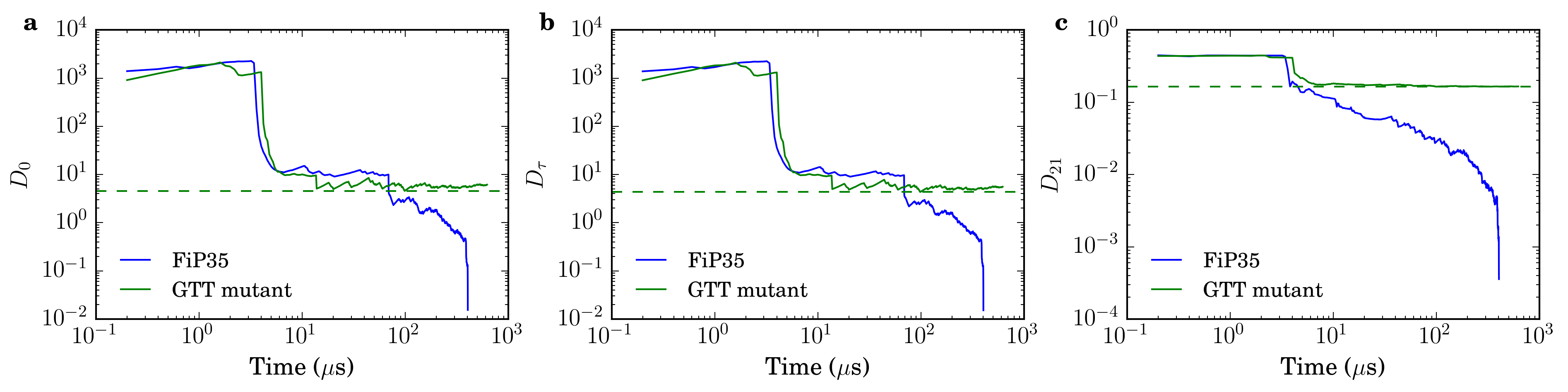}
\caption{Similarity of a) covariance and b) time-lagged covariance matrices for TICs calculated on FiP35 WW domain and GTT mutant simulation sets truncated at varying time-lengths on the full FiP35 simulation set. c) Similarity of the estimated slowest target system (FiP35) TIC using the two slowest transferred TICs from GTT mutant and FiP35 simulation sets. The dashed horizontal green lines show the lowest value attained by any GTT mutant simulation set.}
\label{fig:fip35_on_native}
\end{figure*}

\newpage 

\begin{figure*}[h]
\centering
\includegraphics[width=\textwidth]{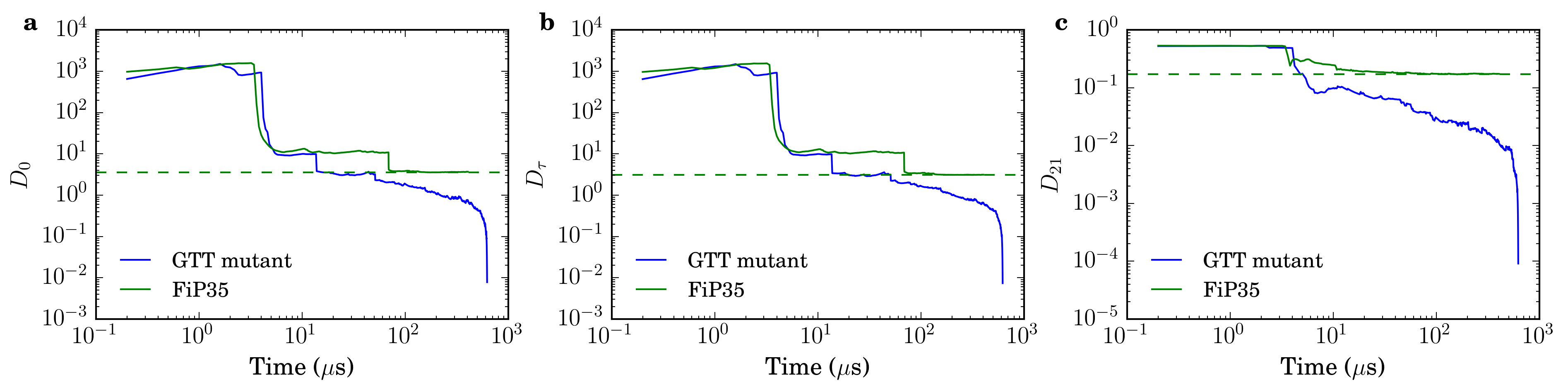}
\caption{Similarity of a) covariance and b) time-lagged covariance matrices for TICs calculated on GTT mutant and FiP35 WW domain simulation sets truncated at varying time-lengths on the full GTT mutant simulation set. c) Similarity of the estimated slowest target system (GTT mutant) TIC using the two slowest transferred TICs from FiP35 and GTT mutant simulation sets. The dashed horizontal green lines show the lowest value attained by any FiP35 simulation set.}
\label{fig:fip35_on_gtt}
\end{figure*}

\newpage

\begin{figure*}[h]
\centering
\includegraphics[width=\textwidth]{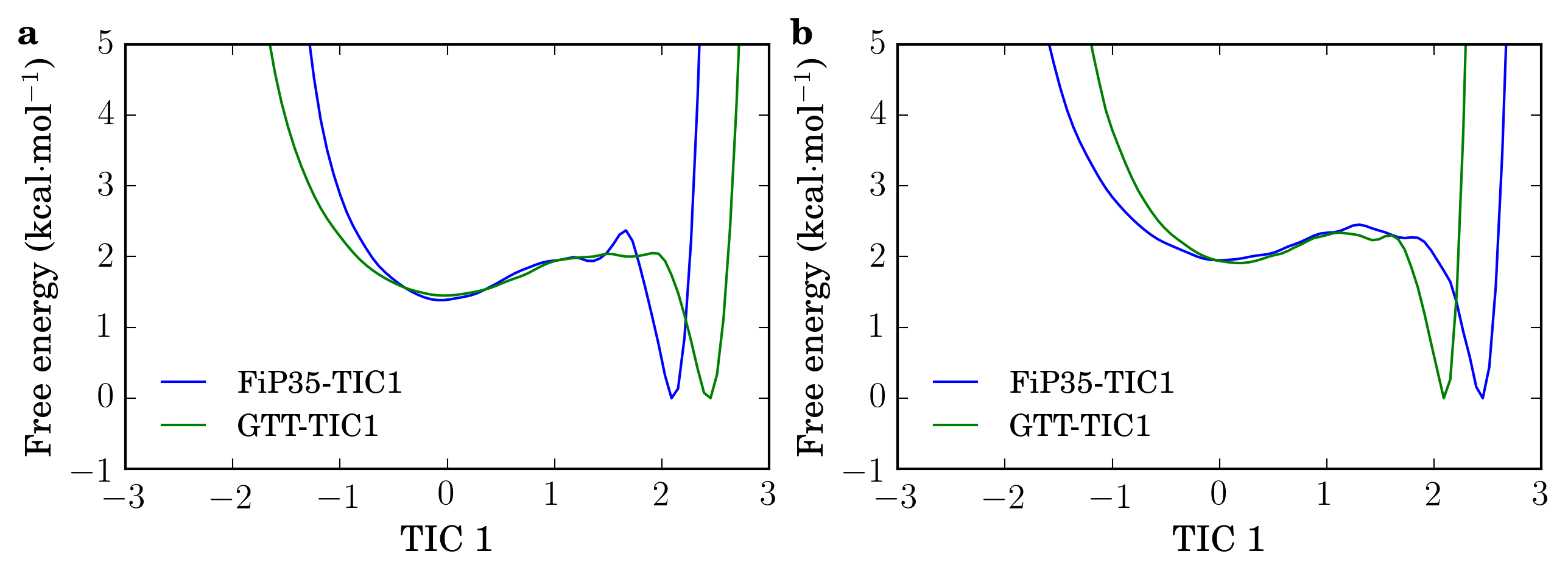}
\caption{Free energy of the a) FiP35 WW domain and b) GTT mutant simulation sets each projected onto both the slowest transferred and native TIC estimated using full datasets. The x-axis represents a projection of the same data onto a different vector for each plot on the same pair of axes.}
\label{fig:fip35_prob}
\end{figure*}

\newpage

\begin{figure*}[h]
\centering
\includegraphics[width=\textwidth]{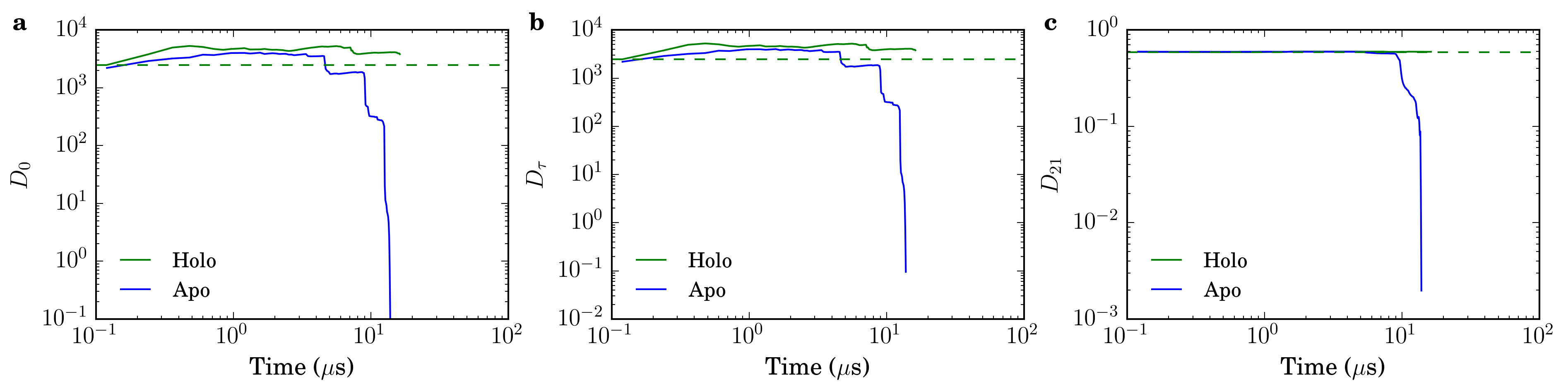}
\caption{Similarity of the a) covariance and b) time-lagged covariance matrices for TICs calculated on apo and holo PKA cBD simulation sets truncated at varying time-lengths on the full apo simulation set. Log-log plots are used to improve visual differentiability of different datasets on the same axes. c) Similarity of the estimated slowest target system (apo) TIC using the two slowest transferred TICs from the apo and holo simulation sets. The dashed horizontal green lines show the value attained by any holo PKA cBD simulation set.}
\label{fig:pka_on_apo}
\end{figure*}

\newpage

\begin{figure*}[h]
\centering
\includegraphics[width=\textwidth]{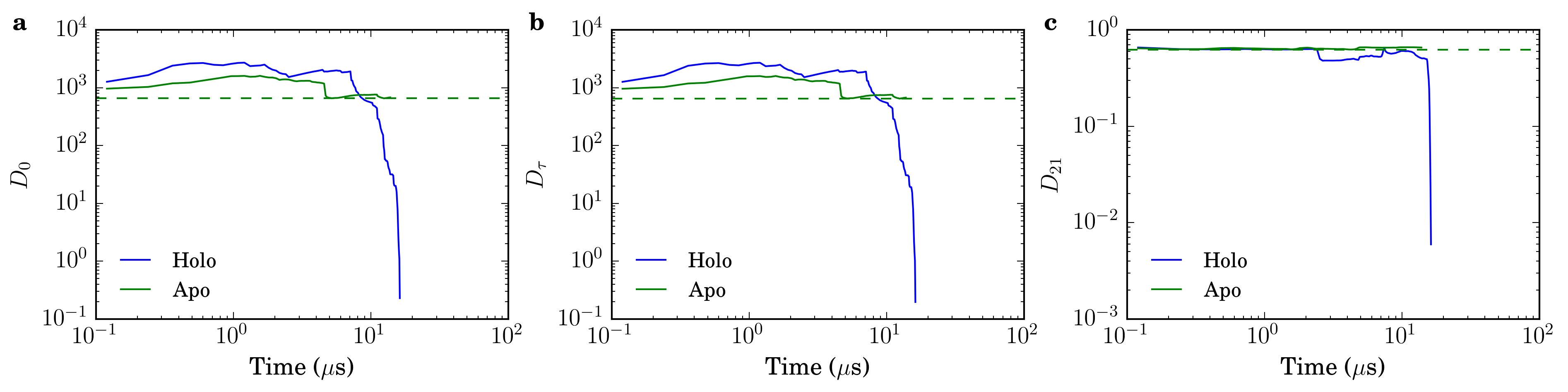}
\caption{Similarity of the a) covariance and b) time-lagged covariance matrices for TICs calculated on holo and apo PKA cBD simulation sets truncated at varying time-lengths on the full holo simulation set. Log-log plots are used to improve visual differentiability of different datasets on the same axes. c) Similarity of the estimated slowest target system (holo) TIC using the two slowest transferred TICs from the holo and apo simulation sets. The dashed horizontal green lines show the value attained by any apo PKA cBD simulation set.}
\label{fig:pka_on_holo}
\end{figure*}

\newpage

\begin{figure*}[h]
\centering
\includegraphics[width=\textwidth]{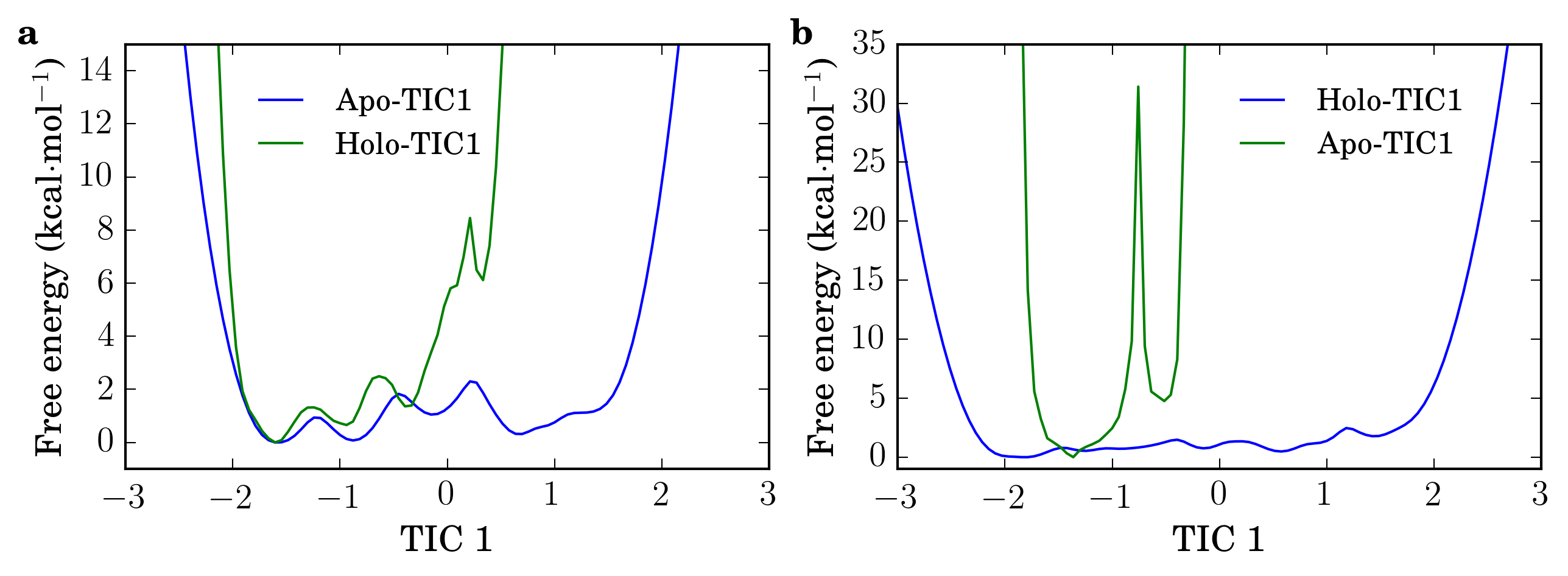}
\caption{Free energy of the a) apo and b) holo PKA cBD simulation sets each projected onto both the slowest transferred and native TIC estimated usin full datasets. The x-axis represents a projection of the same data onto a different vector for each plot on the same pair of axes.}
\label{fig:pka_prob}
\end{figure*}

\newpage

\begin{figure*}[h]
\centering
\includegraphics[width=\textwidth]{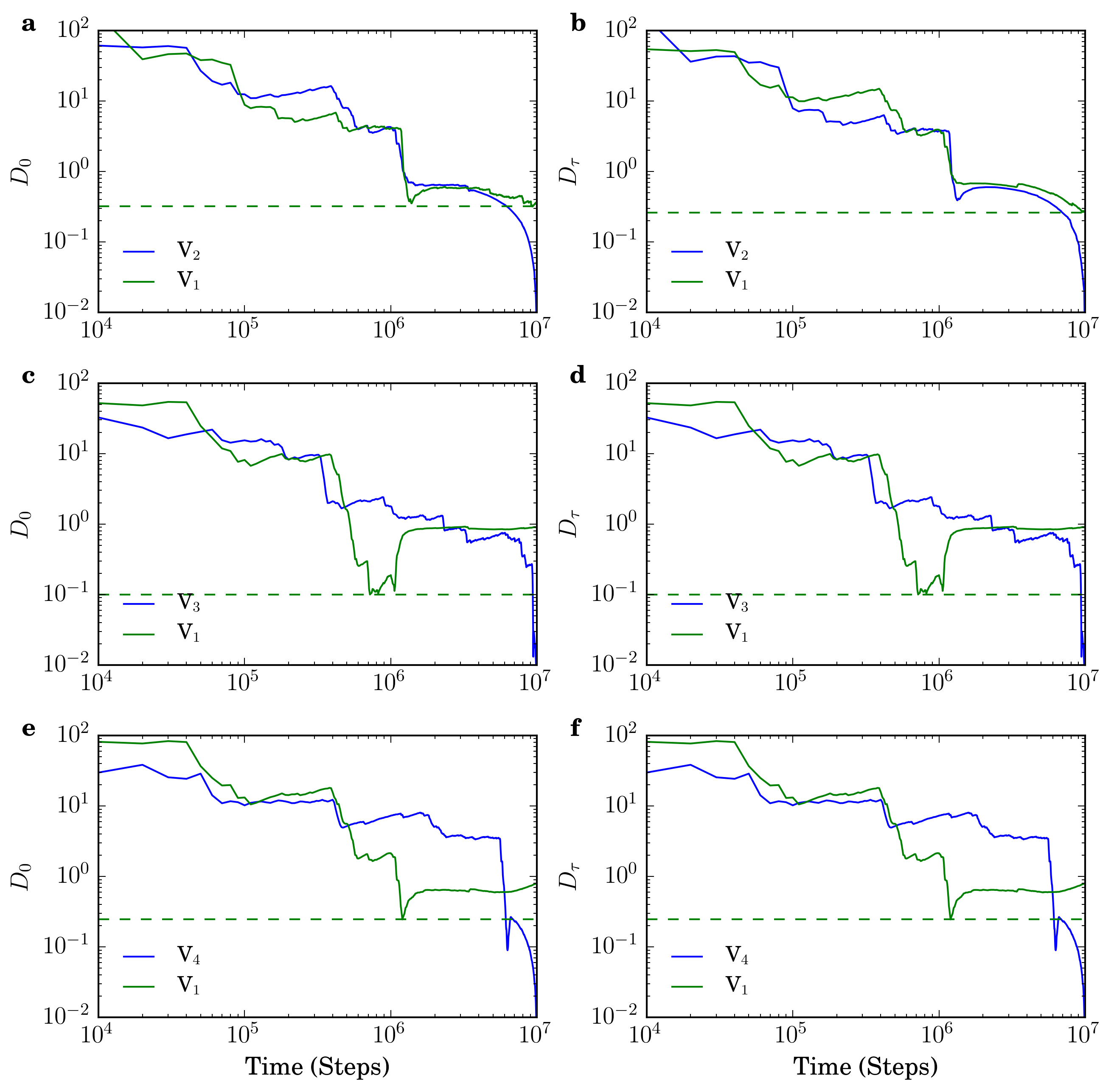}
\caption{Similarity of the a),c),e) covariance and b),d),f) time-lagged covariance matrices for TICs calculated on the V$_{1}$-V$_{4}$ simulation sets truncated at varying time-lengths on the full V$_{2}$-V$_{4}$ simulation sets. V$_{1}$ TICs transferred to a)-b) V$_{2}$, c)-d) V$_{3}$, and e)-f) V$_{4}$. Log-log plots are used to improve visual differentiability of different datasets on the same axes. The dashed horizontal lines show the lowest value attained by any V$_{1}$ simulation set.}
\label{fig:bd_on_v2-4}
\end{figure*}

\newpage

\begin{figure*}[h]
\centering
\includegraphics[width=\textwidth]{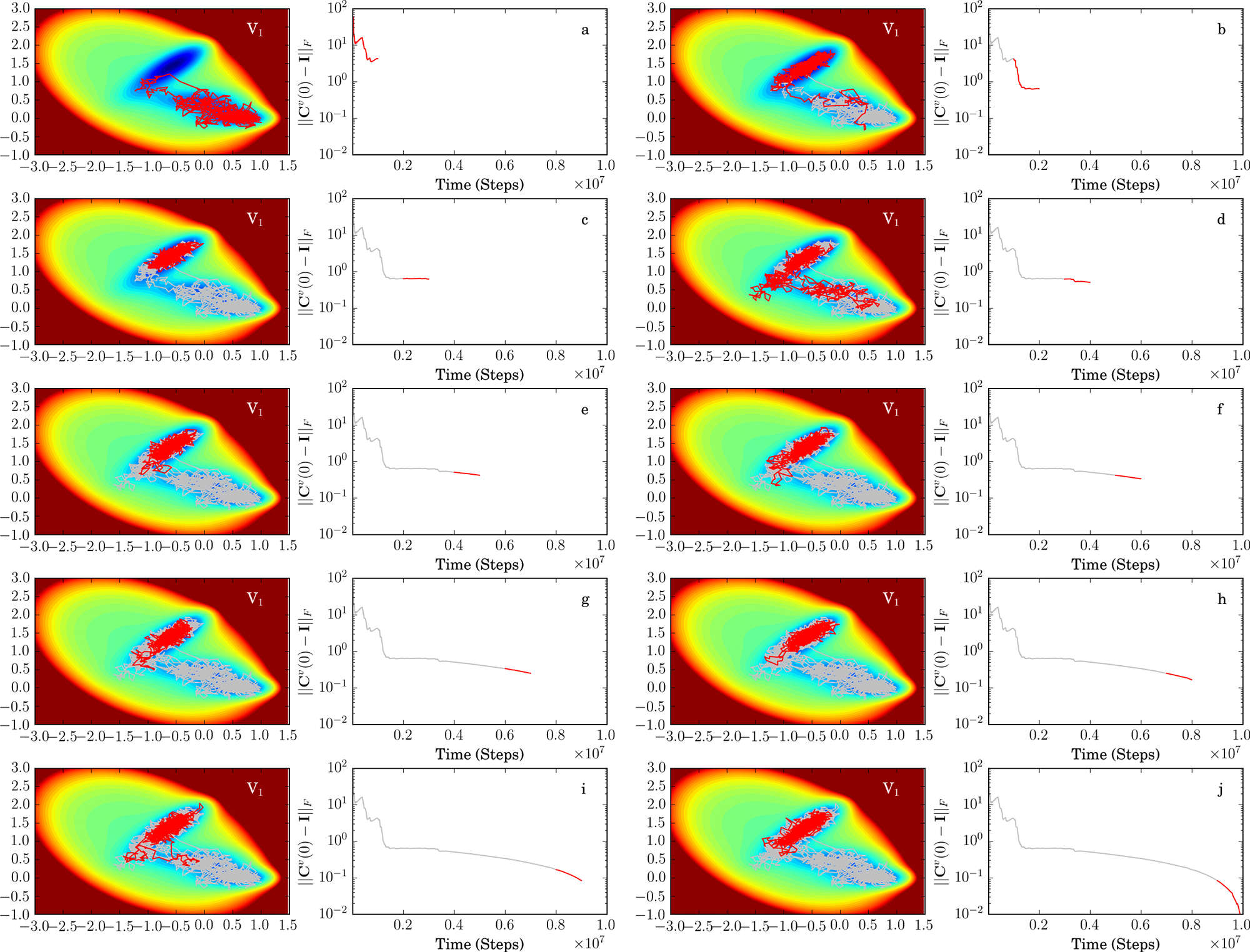}
\caption{Brownian dynamics simulations on the V$_{1}$ potential, where the subfigures a)-j) show sequential sampling intervals, with $D_{0}$ values shown to the right.}
\label{fig:v1_movie}
\end{figure*}
\newpage
\begin{figure*}[h]
\centering
\includegraphics[width=\textwidth]{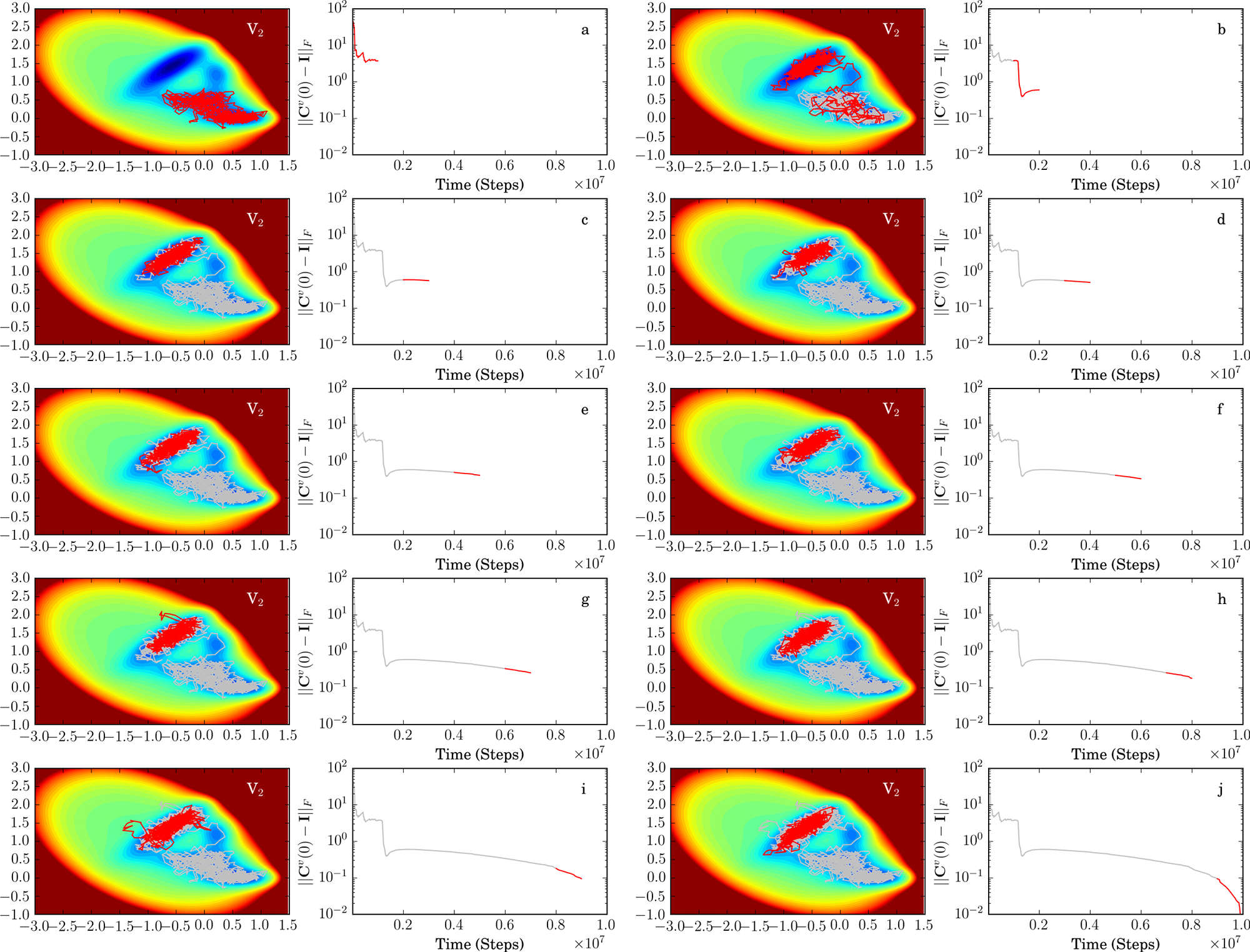}
\caption{Brownian dynamics simulations on the V$_{2}$ potential, where the subfigures a)-j) show sequential sampling intervals, with $D_{0}$ values shown to the right.}
\label{fig:v2_movie}
\end{figure*}
\newpage
\begin{figure*}[h]
\centering
\includegraphics[width=\textwidth]{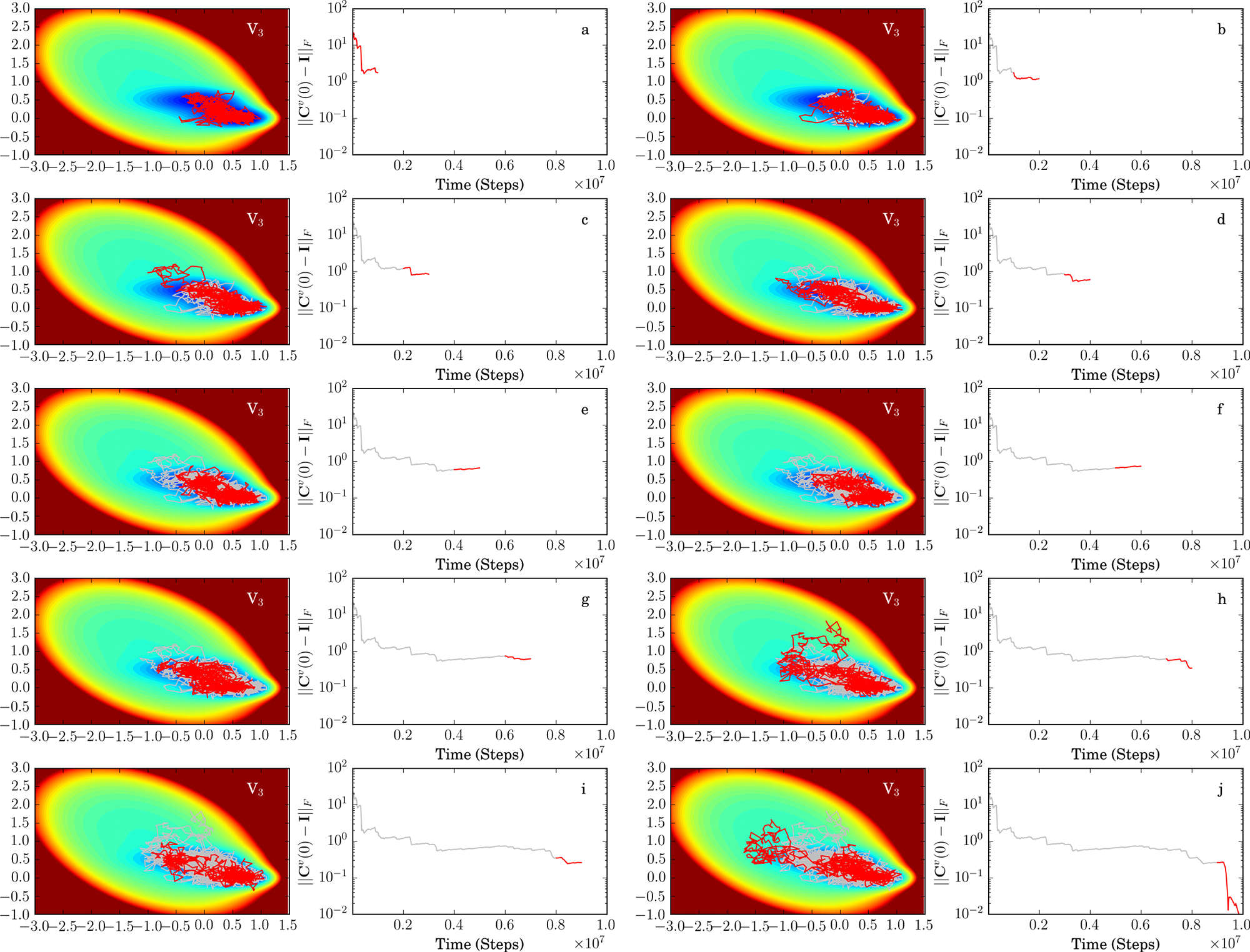}
\caption{Brownian dynamics simulations on the V$_{3}$ potential, where the subfigures a)-j) show sequential sampling intervals, with $D_{0}$ values shown to the right.}
\label{fig:v3_movie}
\end{figure*}
\newpage
\begin{figure*}[h]
\centering
\includegraphics[width=\textwidth]{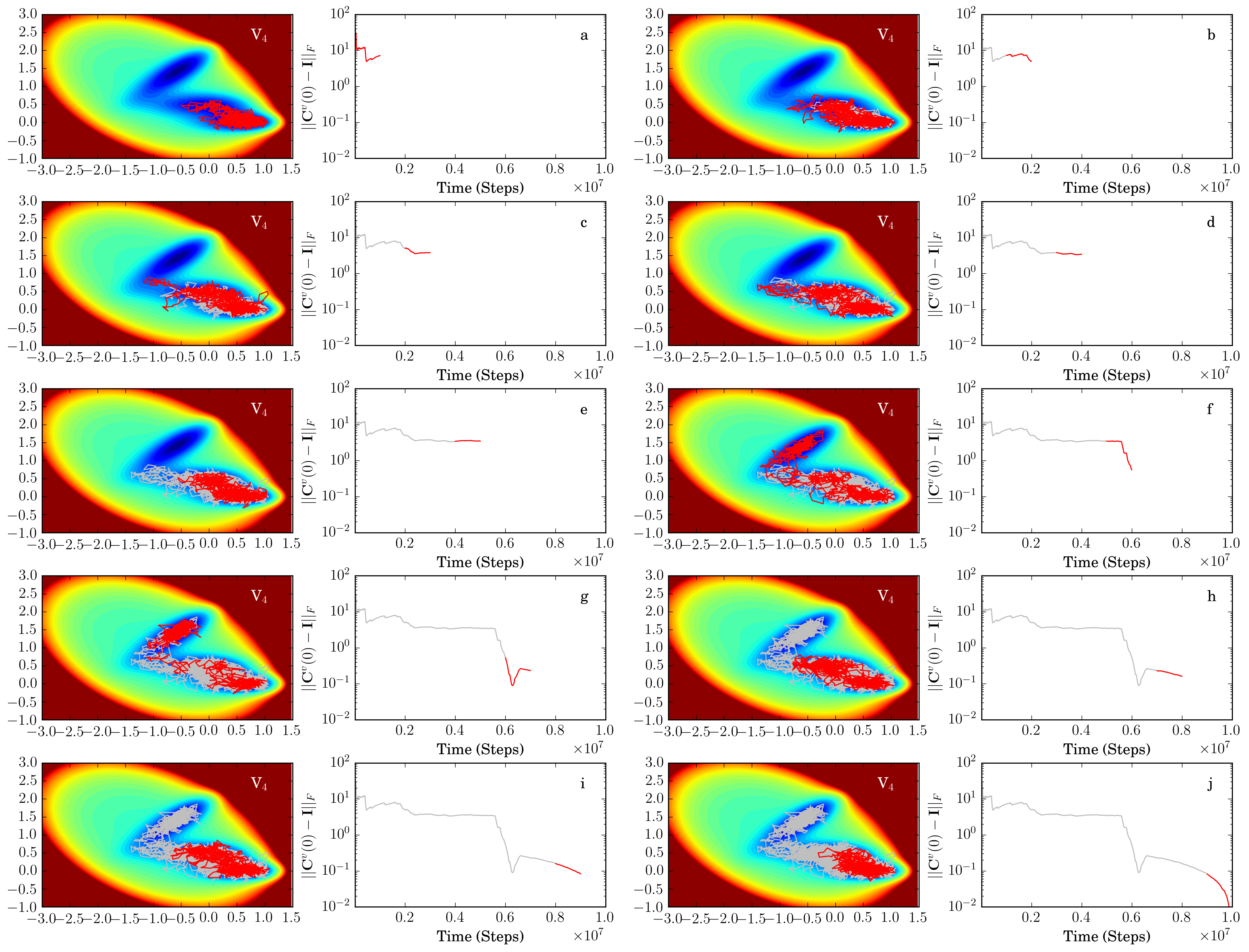}
\caption{Brownian dynamics simulations on the V$_{4}$ potential, where the subfigures a)-j) show sequential sampling intervals, with $D_{0}$ values shown to the right.}
\label{fig:v4_movie}
\end{figure*}
\newpage
\end{widetext}

\end{document}